\documentclass[12pt]{article}
\usepackage{amsfonts}
\usepackage[mathscr]{eucal}
\usepackage[dvips]{color}

\def\theequation{\arabic{section}.\arabic{equation}}
\newcommand{\be}{\begin{equation}}\newcommand{\ee}{\end{equation}}
\newcommand{\bea}{\begin{eqnarray}}
\newcommand{\eea}{\end{eqnarray}}

\newcommand{\lb}{\label}
\newcommand{\p}[1]{(\ref{#1})}

\begin{document}

\begin{titlepage}
\begin{flushright}
ITP-UH-20/09
\end{flushright}

\vspace*{2cm}

\renewcommand{\thefootnote}{\dag}
\begin{center}

{\LARGE\bf New $D(2,1;\alpha)$ Mechanics with Spin Variables}

\vspace{2cm}
\renewcommand{\thefootnote}{\star}

{\large\bf S.~Fedoruk}${}^1$,\,\,\, {\large\bf E.~Ivanov}${}^1$,\,\,\,
{\large\bf
O.~Lechtenfeld}${}^2$ \vspace{1cm}

${}^1${\it Bogoliubov  Laboratory of Theoretical Physics, JINR,}\\
{\it 141980 Dubna, Moscow region, Russia} \\
\vspace{0.1cm}

{\tt fedoruk,eivanov@theor.jinr.ru}\\
\vspace{0.5cm}

${}^2${\it Institut f\"ur Theoretische Physik,
Leibniz Universit\"at Hannover,}\\
{\it Appelstra{\ss}e 2, D-30167 Hannover, Germany} \\
\vspace{0.1cm}

{\tt lechtenf@itp.uni-hannover.de}\\
\vspace{0.3cm} \setcounter{footnote}{0}

\end{center}
\vspace{0.2cm} \vskip 0.6truecm  \nopagebreak

\begin{abstract}
\noindent We elaborate on a novel superconformal mechanics model possessing $D(2,1;\alpha)$
symmetry and involving extra U(2) spin variables. It is the one-particle case of the ${\cal
N}{=}4$ superconformal matrix model recently proposed in {\tt arXiv:0812.4276\,[hep-th]},
and it generalizes to arbitrary $\alpha{\neq}0$ the OSp$(4|2)$ superconformal mechanics of
{\tt arXiv:0905.4951\,[hep-th]}. As in the latter case, the U(2) spin variables  are
described by a Wess-Zumino action and define the first Hopf map $S^3 \rightarrow S^2$ in
the target space. Upon quantization, they represent a fuzzy sphere. We find the classical
and quantum generators of the $D(2,1;\alpha)$ superalgebra and their realization on the
physical states. The super wavefunction encompasses various multiplets of the ${\rm
SU(2)}_R$ and ${\rm SU(2)}_L$ subgroups of $D(2,1;\alpha)$, with fixed isospins. The
conformal potential is determined by the external magnetic field in the Wess-Zumino term,
whose strength is quantized like in the OSp$(4|2)$ case. As a byproduct, we reveal new
invariant subspaces in the enveloping algebra of $D(2,1;\alpha)$ for our quantum
realization.
\end{abstract}

\vspace{1cm}


\newpage

\end{titlepage}

\setcounter{footnote}{0}

\setcounter{equation}0
\section{Introduction}

The interest in various models of ${\cal N}{=}4$ superconformal
mechanics is mainly caused by the possibility of using them for the
description of supergravity black-hole solutions within the
AdS/CFT correspondence, as was first suggested in~\cite{CDKKTP}.

In~\cite{FIL}, we constructed a new ${\cal N}{=}4$ superconformal
matrix model with ${\rm U}(n)$ gauge symmetry. This model is
described by the following harmonic superspace action,
\begin{equation}\label{4N-gau-matrix}
S = -{\textstyle\frac{1}{4(1+\alpha)}}\int \mu_H  {\rm Tr} \left(
\mathscr{X}^{\,-1/\alpha} \, \right) + {\textstyle\frac{1}{2}}\int
\mu^{(-2)}_A \mathcal{V}_0 \widetilde{\mathcal{Z}}{}^+
\mathcal{Z}^+ + {\textstyle\frac{i}{2}}\,c\int \mu^{(-2)}_A \,{\rm
Tr} \,V^{++} \,,
\end{equation}
where $\alpha$ is a real parameter which can take any non-zero
value. The first term in~(\ref{4N-gau-matrix}) is the gauged action
of the ({\bf 1,4,3}) multiplets which are described by hermitian
$(n{\times}n)$-matrix superfields $\mathscr{X}=(\mathscr{X}_a^b)$,
$a,b=1,\ldots ,n$. They are in the adjoint of ${\rm U}(n)$ and are
subject to appropriate gauge-covariant constraints. These
constraints involve the gauge connections which are expressed
through the analytic harmonic gauge superfield $V^{++}(\zeta,u)$
\cite{DI}. The third term in~(\ref{4N-gau-matrix}) is a
Fayet-Iliopoulos (FI) term for $V^{++}$ and the real constant $c$
is its strength. The second term in~(\ref{4N-gau-matrix}) is a
Wess-Zumino~(WZ) action describing $n$ commuting analytic superfields
$\mathcal{Z}^+_a$ which represent off-shell ${\cal N}{=}4$
multiplets of type ({\bf 4,4,0}) and are in the fundamental of ${\rm
U}(n)$. The superfield $\mathcal{V}_0(\zeta,u)$ is a real analytic
gauge prepotential for the ${\rm U}(n)$ singlet ({\bf 1,4,3})
superfield $\mathscr{X}_0 \equiv {\rm Tr} \left( \mathscr{X}
\right)\,$.

After passing to the WZ gauge, eliminating auxiliary degrees of freedom
and fixing a gauge with respect to the residual gauge group, the
model~(\ref{4N-gau-matrix}) involves $n$ bosonic fields $x_a$ which
are the first components of the diagonal superfields $\mathscr{X}^a_a$
(no sum over $a$), $n^2$ fermionic fields $\psi^b_a$ which are the second
components in the $\theta$ expansion of $\mathscr{X}^b_a$, and the lowest
commuting components of the superfields $\mathcal{Z}^+_a\,$. The latter
variables are described by Wess-Zumino-type $d=1$ actions and parametrize
$n$ independent target spheres $S^2\,$. Thus, they may be interpreted as
target harmonic variables. After quantization, they become
a sort of non-dynamical spin variables representing $n$ ``fuzzy'' spheres.

The model~(\ref{4N-gau-matrix}) is invariant under the most general
${\cal N}{=}4$ superconformal symmetry $D(2,1;\alpha)$ (with the more customary
${\rm OSp}(4|2)$ and ${\rm SU}(1,1|2)$ symmetries as particular cases).
It contains two ${\rm{SU}}(2)$ R-symmetry subgroups one of which acts
only on fermions. In the case of $D(2,1;\alpha{=}{-}\frac12)\simeq
{\rm OSp}(4|2)$, this model yields a new ${\cal N}{=}4$ supersymmetric
extension of the ${\rm U(2)}$ spin $A_{n-1}$ Calogero system.

Note that for $\alpha{=}{-}1$ we have $D(2,1;\alpha{=}{-}1)\simeq
{\rm SU(1,1|2)}{\subset\!\!\!\!\!\!\times}{\rm SU(2)}$.
It was argued in~\cite{GT} that the large-$n$ limit of the $n$-particle
${\rm SU(1,1|2)}$ superconformal Calogero model provides a microscopic
description of the extreme Reissner-Nordstr\"om (RN) black hole in the
near-horizon limit. This hypothesis is based on the assertion that for
a large number of particles and in a limit when all coordinates of the
Calogero model, except for one, are treated as ``small'', the Calogero model
reduces to the conformal mechanics for this ``allocated''  coordinate.

For all values of $\alpha{\neq}{-}1/2\,$, the actions~(\ref{4N-gau-matrix})
yield non-trivial conformal sigma models in the bosonic limit.
Therefore, the model~(\ref{4N-gau-matrix}) can hardly be utilized to describe
a single black hole along the lines of~\cite{GT}.
Yet, it may be relevant to the multi-black-hole system, since
the corresponding moduli spaces of $n$ black holes in four- and five-dimensional supergravities
are known to be described by sigma-model-type multi-black-hole
quantum mechanics~\cite{MS}. They become flat precisely in the case
of ${\rm{OSp}}(4|2)$ superconformal symmetry, i.e.~at $\alpha{=}{-}1/2$.

Note that the construction of a self-consistent $n$-body generalization
of black-hole quantum mechanics is a rather complicated problem~\cite{MS}
beyond the one- and two-body cases. In order to have a normalizable ground
state in the latter cases, one should apply a proper time redefinition,
just as in conformal quantum mechanics~\cite{AFF}. If the general
multi-black-hole quantum mechanics amounts to supersymmetric Calogero models,
one can employ the powerful machinery developed for integrable
super-Calogero systems (see e.g.~\cite{FM,Wyl,Vas,GLP}).

In the present paper we investigate the $n{=}1$ case of the
model~(\ref{4N-gau-matrix}), which describes the center-of-mass
motion in the general super-Calogero model and, therefore,
corresponds to a single black hole. The special case of
$\alpha{=}{-}1/2$, both on classical and quantum levels, was
considered in detail in~\cite{FIL-09}. Here, we extend this
consideration to all non-zero values of~$\alpha$.\footnote{Another
view of the $D(2,1;\alpha)$ superconformal mechanics models with spin variables
(based on an $su(2)$ Hamiltonian reduction at the classical component level)
was presented in~\cite{KL-09}.} We hope that an exhaustive
understanding of the $n{=}1$ case will be helpful for attacking the
quantum $D(2,1;\alpha)$ model for arbitrary values of~$n$.

We use the standard notations of ${\cal N}{=}4, d{=}1$ supersymmetric
theories, following~\cite{IKL,IL} and~\cite{FIL-09}.

\setcounter{equation}0
\section{Superfield setup}

The one-particle limit of the model~(\ref{4N-gau-matrix}) involves
superfields corresponding to three off-shell ${\cal N}{=}4$ supermultiplets: ({\bf i}) the
``radial'' multiplet ({\bf 1,4,3}); ({\bf ii}) the Wess-Zumino (``isospin'') multiplet
({\bf 4,4,0}); and ({\bf iii}) the gauge (``topological'') multiplet. The
total action has the form
\begin{equation}\label{4N-gau}
S =S_{\mathscr{X}} + S_{FI} + S_{WZ}\,.
\end{equation}

The first term in~(\ref{4N-gau}) is the standard free action of the ({\bf 1,4,3}) multiplet
($\alpha\neq 0$)
\begin{equation}\label{4N-X}
S_{\mathscr{X}} =-\textstyle{\frac{1}{4(1+\alpha)}}\,\displaystyle{\int} \mu_H \,
\mathscr{X}^{\,-1/\alpha} \,,
\end{equation}
where the even real superfield $\mathscr{X}$ is subjected to the constraints
\begin{equation}  \label{cons-X-g-V}
D^{++} \,\mathscr{X}=0\,,
\end{equation}
\begin{equation}  \label{cons-X-g}
D^{+}{D}^{-} \,\mathscr{X}=0\,,\qquad
    \bar D^{+}\bar D^{-}\, \mathscr{X}=0\,,\qquad
    (D^{+}\bar D^{-} +\bar D^{+}D^{-})\, \mathscr{X}=0\,.
\end{equation}
The set of conditions \p{cons-X-g-V} and~(\ref{cons-X-g}) is equivalent to the standard
constraints $D^iD_i \,\mathscr{X}=0$, $\bar D_i\bar D^i \,\mathscr{X}=0$, $[D^i,\bar D_i]\,
\mathscr{X}=0$ for the superfield $\mathscr{X}$ living in the
``central basis ${\cal N}{=}4$ superspace'' parametrized
by the coordinates $\theta_i$, $\bar\theta^i$ and $t$.

Note that the action~(\ref{4N-X}) is in fact non-singular at $\alpha=-1\,$.
Indeed, making use of the fact that
${\int} \mu_H \, \mathscr{X}$ is an integral of total derivative, we cast the
action~(\ref{4N-X}) in the equivalent form
$$
S_{\mathscr{X}} =-\textstyle{\frac{1}{4(1+\alpha)}}\,\displaystyle{\int} \mu_H \,
\left(\mathscr{X}^{\,-1/\alpha} -\mathscr{X}\right)\,.
$$
Thus in the limit $\alpha=-1$ we obtain the standard action
\begin{equation}\label{4N-X0}
S_{\mathscr{X}} \Big|_{\alpha=-1} =-\textstyle{\frac{1}{4}}\,\displaystyle{\int} \mu_H \,
\mathscr{X} \,\ln\! \mathscr{X} \,,
\end{equation}
The action~(\ref{4N-X}) is not defined at $\alpha{=}0$, and this special case needs a separate
analysis (see Section 5). In what follows we always assume that $\alpha \neq 0\,$.

The second term in~(\ref{4N-gau}) is FI term
\begin{equation}\label{4N-FI}
S_{FI} ={\textstyle\frac{i}{2}}\,c\int \mu^{(-2)}_A \,V^{++}
\end{equation}
for the gauge supermultiplet. The even analytic
gauge superfield $V^{++}(\zeta,u)$, ${D}^{+} \,V^{++}=0$, $ \bar{D}^{+}\,
V^{++}=0\,,$ is subjected to the gauge transformations
\begin{equation} \label{tran4-V}
V^{++}{}' = V^{++} - D^{++}\lambda, \quad \lambda = \lambda(\zeta,
u)\,,
\end{equation}
which are capable to gauge away, {\it locally}, all the components from
$V^{++}$. However, the latter contains a component which cannot be gauged away {\it globally}.
This is the reason why this $d=1$ supermultiplet was called ``topological'' in \cite{DI}.

Last term in~(\ref{4N-gau}) is Wess-Zumino (WZ) term
\begin{equation}\label{4N-WZ}
S_{WZ} = {\textstyle\frac{1}{2}}\,b\int \mu^{(-2)}_A \, \mathcal{V}\,
\widetilde{\mathcal{Z}}{}^+\, {\mathcal{Z}}^+\, .
\end{equation}
Here, the complex analytic superfield ${\mathcal{Z}}^+,
\widetilde{\mathcal{Z}}^+$ $(D^+
{\mathcal{Z}}^+ = \bar D^+ {\mathcal{Z}}^+ =
0)\,,$ is subjected to the harmonic constraints
\begin{equation}  \label{cons-Ph-g}
\mathscr{D}^{++} \,{\mathcal{Z}}^+\equiv (D^{++} + i\,V^{++})
\,{\mathcal{Z}}^+=0\,,\qquad \mathscr{D}^{++}
\,\widetilde{\mathcal{Z}}{}^+\equiv (D^{++} - i\,V^{++}) \,\widetilde{\mathcal{Z}}{}^+=0
\end{equation}
and describes a gauge-covariantized version of the ${\cal N}{=}4$
multiplet ({\bf
4,4,0}). The relevant gauge transformations
are
\begin{equation} \label{tran4-Phi}
{\mathcal{Z}}^+{}' = e^{i\lambda} {\mathcal{Z}}^+,\qquad
\widetilde{\mathcal{Z}}{}^+{}' =
e^{-i\lambda}\widetilde{\mathcal{Z}}{}^+\,.
\end{equation}
We explicitly included a coupling constant $b$ in~(\ref{4N-WZ}) in order to track
the contribution of WZ term to the full component action. Afterwards, this constant
will be put equal to 1.

The superfield $\mathcal{V}(\zeta,u)$ in \p{4N-WZ} is a real analytic gauge superfield
(${D}^{+} \,\mathcal{V}=\bar{D}^{+}\, \mathcal{V}=0$), which is a prepotential solving the
constraints~(\ref{cons-X-g-V})  and~(\ref{cons-X-g}) for $\mathscr{X}$. It is related to the superfield
$\mathscr{X}$ in the central basis by the harmonic integral transform~\cite{DI1}
\begin{equation}  \label{X0-V0}
\mathscr{X}(t,\theta_i,\bar\theta^i)=\int du \,\mathcal{V} \left(t_A,
\theta^+,
\bar\theta^+, u^\pm \right) \Big|_{\theta^\pm=\theta^i u^\pm_i,\,\,\,
\bar\theta^\pm=\bar\theta^i u^\pm_i}\,.
\end{equation}
The unconstrained analytic prepotential $\mathcal{V}$ possesses its own pregauge
freedom
\begin{equation} \lb{ga-V0}
\delta \mathcal{V} = D^{++}\lambda^{--}\,, \quad \lambda^{--} =
\lambda^{--}(\zeta,
u)\,,
\end{equation}
which can be exploited to show that $\mathcal{V}$ describes just the
multiplet
$({\bf 1, 4, 3})$
(after choosing the appropriate Wess-Zumino gauge) \cite{DI1}.
The coupling to the multiplet $({\bf 1, 4, 3})$ in \p{4N-WZ} is introduced
for
ensuring superconformal invariance.
We shall see that, upon passing to components, it gives rise to non-trivial
interactions for the physical fields.
The invariance of \p{4N-WZ} under \p{ga-V0} is ensured by the constraints
\p{cons-Ph-g}.

Besides the gauge ${\rm U}(1)$ symmetry \p{tran4-V}, \p{tran4-Phi} and
pregauge
symmetry \p{ga-V0},
the action~(\ref{4N-gau}) respects the rigid
${\cal N}{=}4$ superconformal symmetry $D(2,1;\alpha)\,$.
All
superconformal transformations are
contained in the closure of the supertranslations and superconformal boosts.

Invariance of the action~(\ref{4N-gau}) under the supertranslations
($\bar\varepsilon^i=\overline{(\varepsilon_i)}$)
$$
\delta t =i(\theta_k\bar\varepsilon^k -\varepsilon_k\bar\theta^k),\qquad
\delta
\theta_k=\varepsilon_k, \qquad \delta \bar\theta^k=\bar\varepsilon^k
$$
is automatic because we use the ${\cal N}{=}4$ superfield approach.

The coordinate realization of the superconformal boosts of $D(2,1;\alpha)$ \cite{IL,DI} is
as follows ($\bar\eta^i=\overline{(\eta_i)}$):
\begin{equation}  \label{sc-coor-c-b}
\delta^\prime t=i t(\theta_k \bar\eta^k + \bar\theta^k\eta_k) + (1+\alpha)\,\theta_i
\bar\theta^i (\theta_k \bar\eta^k + \bar\theta^k\eta_k)\,,
\end{equation}
\begin{equation}  \label{sc-coor-c-f1}
\delta^\prime \theta_i= \eta_i t -2i\alpha \,\theta_i(\theta_k \bar\eta^k) +
2i(1+\alpha)\,\theta_i (\bar\theta^k\eta_k)- i(1+2\alpha)\,\eta_i (\theta_k
\bar\theta^k)\,,
\end{equation}
\begin{equation}  \label{sc-coor-c-f2}
\delta^\prime \bar\theta^i= \bar\eta^i t -2i\alpha \,\bar\theta^i(\bar\theta^k\eta_k) +
2i(1+\alpha)\,\bar\theta^i (\theta_k \bar\eta^k)+ i(1+2\alpha)\,\bar\eta^i (\theta_k
\bar\theta^k)\,,
\end{equation}
\begin{equation}  \label{sc-coor-bo}
\delta^\prime t_A=\alpha^{-1}\Lambda t_A\,,\qquad \delta^\prime u^+_i= \Lambda^{++}u^-_i\,,
\end{equation}
\begin{equation}  \label{sc-coor-fer}
\delta^\prime \theta^+= \eta^+ t_A +2i(1+\alpha)\eta^- \theta^+\bar\theta^+\,,\qquad
\delta^\prime \bar\theta^+= \bar\eta^+ t_A +2i(1+\alpha)\bar\eta^- \theta^+\bar\theta^+ \,,
\end{equation}
\begin{equation}  \label{sc-me1}
\delta^\prime (dtd^4\theta)=-\alpha^{-1}\,(dtd^4\theta)\,\Lambda_0\,,\qquad \delta^\prime
\mu_H= \mu_H\left(2\Lambda -\alpha^{-1}(1+\alpha)\Lambda_0\right)\,,\qquad \delta^\prime
\mu^{(-2)}_A= 0\,,
\end{equation}
where
\begin{equation}  \label{def-La1}
\Lambda = \tilde\Lambda = 2i\alpha (\bar\eta^-\theta^+ -\eta^-\bar\theta^+ )\,,\quad
\Lambda^{++} = D^{++}\Lambda=2i\alpha (\bar\eta^+\theta^+ - \eta^+\bar\theta^+ )\,, \quad
D^{++}\Lambda^{++}=0\,,
\end{equation}
\begin{equation}  \label{def-La2}
\Lambda_0 = 2\Lambda - D^{--} \Lambda^{++} = 2i\alpha (\theta_k \bar\eta^k +
\bar\theta^k\eta_k)\,, \qquad D^{++}\Lambda_0=0\,.
\end{equation}
Taking the field transformations in the form (here we use the ``passive''
interpretation of them)
\begin{equation}  \label{sc-1n}
\delta^\prime \mathscr{X}= -\Lambda_0\,\mathscr{X}\,,\qquad \delta^\prime
\mathcal{V} =
-2\Lambda\,\mathcal{V}\,,\qquad \delta^\prime {\mathcal{Z}}^+ =
\Lambda\,{\mathcal{Z}}^+\,,\qquad\delta^\prime V^{++} = 0\,,
\end{equation}
it is easy to check the invariance of the action~(\ref{4N-gau}). Note that the
constraints~(\ref{cons-X-g-V}), (\ref{cons-X-g}) and~(\ref{cons-Ph-g}) as well as the
actions~(\ref{4N-X}), (\ref{4N-FI}) and~(\ref{4N-WZ}), are invariant with respect to the
$D(2,1;\alpha)$ transformations with an arbitrary $\alpha{\neq} 0$. It is worth pointing out that the
action~(\ref{4N-WZ}) is superconformally invariant just due to the presence of the analytic
prepotential $\mathcal{V}\,$.

\setcounter{equation}0

\section{Component actions}
\subsection{Action for {\bf (1,4,3)} supermultiplet}

The solution of the constraint~(\ref{cons-X-g-V}), (\ref{cons-X-g}) is as
follows (in the analytic basis):
\begin{eqnarray}
\mathscr{X}&=& x +\theta^- \psi^+ + \bar\theta^- \bar\psi^+ - \theta^+
\psi^- -
\bar\theta^+ \bar\psi^-
  +\theta^-\bar\theta^- N^{++} + \theta^+\bar\theta^+ N^{--} +
(\theta^-\bar\theta^+ +
\theta^+\bar\theta^-) N \nonumber \\
&& + \,\theta^-\theta^+\bar\theta^- \Omega^+ + \bar\theta^-\bar\theta^+\theta^-
\bar\Omega^+
+ \theta^-\bar\theta^-\theta^+\bar\theta^+ D\,. \label{X-WZ}
\end{eqnarray}
Here
\begin{equation}\label{N-WZ}
N^{\pm\pm} = N^{ik}u_i^\pm u_k^\pm \,,\qquad  N = i \dot x - N^{ik}u_i^+
u_k^- \,,
\qquad D = 2\ddot{x} +2i \dot{N}{}^{ik}u_i^+ u_k^- \,,
\end{equation}
\begin{equation}\label{Psi-WZ}
\psi^{\pm} = \psi^{i}u_i^\pm  \,,\qquad \bar\psi{}^{\pm} =
\bar\psi{}^{i}u_i^\pm \,,
\qquad
\Omega^{+} = 2i\dot{\psi}{}^+  \,,\qquad \bar\Omega^{+} =
-2i\dot{\bar\psi}{}^+
\end{equation}
and $x(t_A)$, $N^{ik}= N^{(ik)}(t_A)$, $\psi^{i}(t_A)$,
$\bar\psi_{i}(t_A)=(\overline{\psi^{i}})$ are $d{=}1$ fields.

Inserting~(\ref{X-WZ}) in~(\ref{4N-X})
we obtain
\begin{eqnarray}\label{4N-X-WZ}
S_{\mathscr{X}} &=& \textstyle{\frac{1}{4\alpha^2}}\,\displaystyle{\int} dt\,
x^{-\frac{1}{\alpha}-2}\,\left[ \dot x\dot x -i \left(\bar\psi_k \dot\psi^k
-\dot{\bar\psi}_k \psi^k \right) -{\textstyle\frac{1}{2}}N^{ik}N_{ik}\, \right] \\
&&  -\, \textstyle{\frac{1}{4\alpha^2}}\, (\textstyle{\frac{1}{\alpha}}+2)
\,\displaystyle{\int} dt\, x^{-\frac{1}{\alpha}-3}\, N^{ik} \psi_{(i}\bar\psi_{k)}    - \,
\textstyle{\frac{1}{12\alpha^2}}\,
(\textstyle{\frac{1}{\alpha}}+2)(\textstyle{\frac{1}{\alpha}}+3)\, \displaystyle{\int} dt\,
x^{-\frac{1}{\alpha}-4}\, \psi^{i}\bar\psi^{k} \psi_{(i}\bar\psi_{k)} \,.\nonumber
\end{eqnarray}

In the central basis the $\theta$ expansion~(\ref{X-WZ}) takes the form:
\begin{equation}  \label{sing-X0-WZ}
\mathscr{X}(t,\theta_i,\bar\theta^i)= x + \theta_i\psi^i +
\bar\psi_i\bar\theta^i +
\theta^i\bar\theta^k
N_{ik}+{\textstyle\frac{i}{2}}(\theta)^2\dot{\psi}_i\bar\theta^i
+{\textstyle\frac{i}{2}}(\bar\theta)^2\theta_i\dot{\bar\psi}{}^i +
{\textstyle\frac{1}{4}}(\theta)^2(\bar\theta)^2 \ddot{x}\,,
\end{equation}
where $(\theta)^2\equiv \theta_i\theta^i=-2\theta^+\theta^-$,
$(\bar\theta)^2\equiv
\bar\theta^i\bar\theta_i=2\bar\theta^+\bar\theta^-\,$.
Then, from~(\ref{X0-V0}) we can identify the fields appearing in the WZ
gauge for $\mathcal{V}$ with the fields in \p{sing-X0-WZ}
\begin{equation}  \label{V0-WZ}
\mathcal{V} (t_A, \theta^+, \bar\theta^+, u^\pm) =x(t_A)- 2\,\theta^+
\psi^{i}(t_A)u^-_i  -
2\,\bar\theta^+ \bar\psi^{i}(t_A)u^-_i + 3\,\theta^+ \bar\theta^+
N^{ik}(t_A)u^-_i
u^-_k
\,.
\end{equation}
This expansion will be used to express the action~(\ref{4N-WZ}) in terms
of the
component fields.

\subsection{FI and WZ actions}

Using the ${\rm U}(1)$ gauge freedom~\p{tran4-V}, (\ref{tran4-Phi}) we can
choose WZ
gauge
\begin{equation}  \label{WZ-4N}
V^{++} =-2i\,\theta^{+}
    \bar\theta^{+}A(t_A)\,.
\end{equation}
Then
\begin{equation}
S_{FI} = c \int dt \,A\,.
\end{equation}

The solution of the constraint~(\ref{cons-Ph-g}) in WZ gauge~(\ref{WZ-4N}) is
$$
{\mathcal{Z}}^+ = z^{i}u_i^+ + \theta^+ \varphi + \bar\theta^+ \phi + 2i\,
\theta^+
\bar\theta^+\nabla_{t_A}z^{i}u_i^-
\,,\qquad \widetilde{\mathcal{Z}}{}^+ = \bar z_{i}u^{+i} + \theta^+ \bar\phi -
\bar\theta^+
\bar\varphi + 2i\, \theta^+ \bar\theta^+ \nabla_{t_A}\bar z_{i}u^{-i}
$$
where
\begin{equation}\label{cov-Z}
\nabla z^k=\dot z^k + i A \, z^k\,, \qquad\nabla \bar z_k=\dot{\bar z}_k
-i A\,\bar
z_k\,.
\end{equation}
In~(\ref{cov-Z}), $z^{i}(t_A)$ and $\varphi(t_A)$, $\phi(t_A)$  are
$d{=}1$ fields,
bosonic and fermionic,
respectively. The fields $z^i$ form a complex doublet of the R-symmetry ${\rm
SU}(2)$ group, while the fermionic fields
are singlets of the latter. Another (``mirror'') R-symmetry ${\rm SU}(2)$
is not
manifest in the present approach: the bosonic
fields are its singlets, while the fermionic fields form a doublet with
respect to it.

Inserting expressions~(\ref{cov-Z}) and (\ref{V0-WZ}) in the
action~(\ref{4N-WZ}) and performing integration over $\theta\,$s and harmonics there,
we obtain a component form of the WZ action
\begin{eqnarray}\label{4N-WZ-WZ}
S_{WZ} &=& {\textstyle\frac{i}{2}}\,b\int dt \,\left(\bar z_k \nabla z^k -
\nabla \bar z_k
\, z^k \right)x - {\textstyle\frac{1}{2}}\,b\int dt \,N^{ik}\bar z_i z_k
\\
\nonumber && \, + {\textstyle\frac{1}{2}}\,b\int dt \, \Big[\psi^k
\left(\bar\varphi\,
z_k+
\bar z_k\phi\right) + \bar\psi^k \left(\bar{\phi\,\,}\! z_k- \bar
z_k\varphi\right) - x
\left(\bar{\phi\,\,}\! \phi+ \bar\varphi\,\varphi\right)\Big] \, .
\end{eqnarray}
The fermionic fields $\phi, \varphi$ are auxiliary. The action is
invariant under
the residual local ${\rm U}(1)$ transformations
\begin{equation} \label{res-ga}
A' = A - \dot{\lambda}_0\,, \quad z^i{}' = e^{i\lambda_0}z^i\,, \;
\bar{z}_i{}' =
e^{-i\lambda_0}\bar{z}_i
\end{equation}
(and similar phase transformations of the fermionic fields).

The total component action is a sum of \p{4N-X-WZ}, \p{4N-FI-WZ} and
\p{4N-WZ-WZ}.
Eliminating the auxiliary fields $N^{ik}$, $\phi$,  $\bar\phi$, $\varphi$,
$\bar\varphi$,
from this sum by their algebraic equations of motion,
\begin{equation}\label{4N-eq-N}
N_{ik} = -2b\alpha^{2} x^{\frac{1}{\alpha}+2} z_{(i} \bar z_{k)}-
(\textstyle{\frac{1}{\alpha}}+2)\, x^{-1}\, \psi_{(i}\bar\psi_{k)} \,,
\end{equation}
\begin{equation}\label{4N-eq-phi}
\phi =-\frac{\bar\psi^{k} z_{k}}{x} \,, \quad \bar\phi =\frac{\psi^{k} \bar z_{k}}{x} \,,
\quad \varphi =-\frac{\psi^{k} z_{k}}{x}\,, \quad \bar\varphi =-\frac{\bar\psi^{k} \bar
z_{k}}{x}\,,
\end{equation}
and making the redefinition
\begin{equation}\label{4N-nZ}
x^{\prime} = x^{-\frac{1}{2\alpha}} \,, \qquad \psi_{k}^{\prime} =
-{\textstyle\frac{1}{2\alpha}}\,x^{-\frac{1}{2\alpha}-1} \psi_{k}\,, \qquad z^\prime{}^i =
x^{1/2}\,z^i\,,
\end{equation}
we obtain the on-shell form of the action~(\ref{4N-gau}) in WZ gauge (we
omitted the primes on $x$, $\psi$ and $z$)
\begin{eqnarray}
S &=& S_b+ S_f\,, \label{4N-ph}\\
S_b &=&  \int dt \,\Big[\dot x\dot x  + {\textstyle\frac{i}{2}}\,b \left(\bar z_k \dot z^k -
\dot{\bar z}_k z^k\right)-\frac{b^2\alpha^2(\bar z_k
z^{k})^2}{4x^2} -A \left(b\bar z_k z^{k} -c \right) \Big] \,,\label{bose}\\
S_f &=&  -i \!\int\! dt \left( \bar\psi_k \dot\psi^k -\dot{\bar\psi}_k \psi^k \right) +
2b\alpha  \!\int\! dt \, \frac{\psi^{i}\bar\psi^{k} z_{(i} \bar z_{k)}}{x^2} +
\textstyle{\frac{2}{3}}\,(1+2\alpha) \!\displaystyle{\int}\! dt\,
\frac{\psi^{i}\bar\psi^{k} \psi_{(i}\bar\psi_{k)}}{x^2}  . \label{fermi}
\end{eqnarray}

It is still invariant under the gauge transformations \p{res-ga}. The $d{=}1$ gauge connection
$A(t)$ in \p{bose} is the Lagrange multiplier for the constraint
\begin{equation}\label{con}
\bar z_k z^{k}=c\,.
\end{equation}
This constraint implies $c>0$.
After varying with respect to $A$, the action \p{4N-ph} is gauge invariant
only with
taking into account
the constraint \p{con} which is gauge invariant by itself.
The constant $b$ in \p{bose}, \p{fermi} marks the contributions of the superfield WZ term to
the physical component action. It can be eliminated by a proper rescaling of the variables $z^i, \bar z_i$,
so hereafter we choose $b=1\,$.

It is convenient to fully fix the residual gauge freedom by choosing the phases of $z^1$
and $z^2$ opposite to each other. In this gauge, the constraint \p{con} is solved by
\begin{equation}\label{ga-u1}
z^{1} = \kappa \cos{\textstyle\frac{\gamma}{2}}\,e^{i\beta/2}\,,\qquad z^{2}= \kappa
\sin{\textstyle\frac{\gamma}{2}}\,e^{-i\beta/2}\,,\qquad \kappa^2=c\,.
\end{equation}
In terms of the newly introduced  fields the bosonic action~(\ref{bose}) takes the form\footnote{
The fermionic action~(\ref{fermi}) can also be rewritten in terms of $\beta$ and $\gamma\,$,
like its $\alpha{=}-1/2$ prototype~\cite{FIL-09}.}
\begin{equation}\label{bose1}
S_b =  \int dt \,\Big[\dot x\dot x  -\frac{\alpha^2c^2}{4\,x^2} - \frac{c}{2}\,
\cos\gamma\, \dot \beta  \Big] \,.
\end{equation}
As argued in Section 5, this action can be relevant to describing
some particular orbits near horizon of the extreme $D{=}5$ black
holes. The spinor $z^{k}$ provides a parametrization of the angular part of the set
of the horizon coordinates.

Unconstrained fields in the action~(\ref{4N-ph}), three bosons $x$, $\gamma$, $\beta$ and
four fermions $\psi^k$, $\bar\psi_k$, constitute some on-shell
supermultiplet with three bosonic and four fermionic fields.
As opposed to the off-shell ({\bf 3,4,1}) supermultiplet considered
in~\cite{IKL,IKLech,IKLecht} the action~(\ref{bose}) contains ``true'' kinetic term only
for one bosonic component $x$ which also possesses the conformal potential, whereas two other
fields parametrizing the coset SU(2)$_R$/U(1)$_R$ are described by a WZ term.
Taken separately, the WZ term provides an example of Chern-Simons mechanics \cite{FJ,RPS,HTown,Poly,IM}.
The variables $\gamma(t)$ and $\beta(t)$ (or  $z^{k}$ and $\bar z_k$ in the manifestly SU(2)
covariant formulation) become (iso)spin degrees of freedom (target SU(2) harmonics) upon quantization. The
realization of the $D(2,1;\alpha)$ superconformal transformations on these fields will be given
in the next Section.

It should be stressed that the considered model realizes a new mechanism of generating
conformal potential $\sim 1/x^2$ for the field $x(t)$. Before eliminating auxiliary fields,
the component action contains no explicit term of this kind. It arises as a result of
varying with respect to the Lagrange multiplier $A(t)$ and making use of the arising
constraint \p{con}. As we shall see, in quantum theory this new mechanism entails a
quantization of the constant $c\,$.

The naive inspection of the bosonic action~(\ref{bose}) could lead to the conclusion that
angular variables completely decouple from a radial variable, and, hence, are superfluous.
Moreover, the classical dynamics associated with the WZ term in \p{bose} is trivial.
However, like in other Chern-Simons-type theories, this term has a non-trivial impact on
the quantum properties of the model. Indeed, as we shall see in the quantum case, owing to
the non-trivial geometry of the angular space the quantum state vectors necessarily carry
quantum numbers of the ${\rm SU(2)}$ spin. Though in the bosonic limit this symmetry is
purely internal (it commutes with the $d=1$ conformal group ${\rm SL}(2,R)$), the presence
of angular variables leads to the property that the wave function encompasses non-trivial
${\rm SU(2)}$ multiplets.\footnote{In the bosonic case, in accord with the general concept
of separating variables,  one can postulate that the wave function is a product of the
chargeless conformal mechanics wave function by the lowest Landau level wave function
associated with the SU(2) WZ term. No such a separation is possible in the generic
superconformal case due to the presence of fermions interacting with both types of bosonic
variables.} In the supersymmetry case, when the full action~(\ref{4N-ph}) is considered,
the situation becomes even more involved. Now this ${\rm SU(2)}$ symmetry in addition acts
on fermions in parallel with the second ${\rm SU(2)}$ R-symmetry which from the very
beginning is realized only on fermionic fields, and these either ${\rm SU(2)}$ are an
essential part of the superconformal group. Examining the action~(\ref{4N-ph}), we were not
able to find any change of variables which would decouple the angular variables from other
ones. Actually, we already observed the same phenomenon in the particular ${\rm OSp}(4|2)$
case \cite{FIL-09}. Now we see that it persists at any choice of the parameter $\alpha$ in
$D(2,1;\alpha)\,$. Even at the classical level, the WZ term yields, e.g., a non-trivial additional
contribution to the fermionic equations of motion (coming from the term proportional to $b$
in \p{fermi}). Although in the $\beta, \gamma$ parametrization both $\gamma$ and $\dot\beta$
can be expressed through fermions and some integration constant by their classical equations of motion,
an essential trace of the WZ couplings still remains in the equations of motion for fermions,
producing a mass term for them and modifying the coefficients before the third order
terms.\footnote{We thank S.~Krivonos for a discussion on this issue.}
The Hamiltonian, ${\cal N}{=}4$ supercharges and other $D(2,1;\alpha)\,$ generators also involve important new pieces caused
by the WZ term and additional fermionic couplings associated with it (see below).

\subsection{${\cal N}{=}4$ superconformal symmetry in WZ gauge}

The transformations and their generators look most transparent
in terms of the SU(2) doublet quantities $z^{k}$ and $\bar z_{k}$.

To determine the superconformal transformations of component fields, we
should know the
appropriate compensating gauge transformations needed to preserve the WZ
gauge~(\ref{WZ-4N}).
For supertranslations and superconformal boosts the parameter of the compensating
gauge transformations is as follows
\begin{equation}\label{tran-g-WZ}
\lambda=2i\left[ (\theta^+\bar\varepsilon^- -\bar\theta^+\varepsilon^-) + t_A\,
(\theta^+\bar\eta^- -\bar\theta^+\eta^-) \right]A
\end{equation}
where \be \varepsilon^- := \varepsilon^iu^-_i\,, \quad \eta^- := \eta^iu^-_i\,. \ee Taking
this into account, we obtain the relevant infinitesimal $D(2,1;\alpha)$ transformations which
leave the action~(\ref{4N-ph}) invariant (as in~(\ref{4N-ph}) we omit `primes' on
the newly introduced variables):
\begin{equation}\label{2str-X}
\delta x=-\omega_i\psi^i+ \bar\omega^i\bar\psi_i\,,
\end{equation}
\begin{equation}\label{2str-Psi}
\delta \psi^i=i\bar\eta^i x -i \bar\omega^i\dot x -\alpha\frac{\bar\omega_k z^{(i} \bar
z^{k)}}{x} -(1+2\alpha)\frac{\bar\omega_k \psi^{k} \bar\psi^{i} +\omega_k \psi^{k}
\psi^{i}}{x}\,,
\end{equation}
\begin{equation}\label{2str-Psib}
\delta \bar\psi_i=-i\eta_i x +i \omega_i\dot x -\alpha\frac{\omega^k z_{(i} \bar z_{k)}}{x}
+(1+2\alpha)\frac{\omega^k \bar\psi_{k} \psi_{i} +\bar\omega^k \bar\psi_{k}
\bar\psi_{i}}{x}\,,
\end{equation}
\begin{equation}\label{2str-Z}
\delta z^i=-2\alpha\,\frac{\omega^{(i}\psi^{k)} +\bar\omega^{(i}\bar\psi^{k)}}{x}\,z_k\,,
\qquad \delta \bar z_i=2\alpha\,\frac{\omega_{(i}\psi_{k)}
+\bar\omega_{(i}\bar\psi_{k)}}{x}\,\bar z^k\,,
\end{equation}
\begin{equation}\label{2str-A}
\delta A=0\,,
\end{equation}
where $\omega_i = \varepsilon_i + t\, \eta_i\, $ and $\bar\omega^i =
\bar\varepsilon^i + t\, \bar\eta^i\, $. Note that the closure of $d{=}1$
Poincar\`e supersymmetry  transformations is a sum of the time
translations and residual ${\rm U}(1)$ gauge transformation with a
field-dependent parameter. Such a sum turns out to vanish for the
$d{=}1$ gauge field $A$. In Appendix this specifically $d{=}1$
phenomenon is expounded on a simple example of toy ${\cal N}{=}2$
supersymmetric model.

Now, using the N\"other procedure, we can directly find the classical generators of the
supertranslations
\begin{equation}\label{Q-cl}
{Q}^i =p\, \psi^i+ 2i\alpha\frac{z^{(i} \bar z^{k)}\psi_k}{x}+ i(1+2\alpha)\frac{\psi_{k}
\psi^{k}\bar\psi^i}{x}\, ,
\end{equation}
\begin{equation}\label{Qb-cl}
\bar{Q}_i=p\, \bar\psi_i- 2i\alpha\frac{z_{(i} \bar z_{k)}\bar\psi^k}{x}+
i(1+2\alpha)\frac{\bar\psi^{k} \bar\psi_{k}\psi_i}{x}\,,
\end{equation}
where $ p\equiv 2\dot x$, as well as of the superconformal boosts:
\begin{equation}\label{S-cl}
{S}^i =-2\,x \psi^i + t\,{Q}^i,\qquad \bar{S}_i=-2\,x
\bar\psi_i+t\,\bar{Q}_i\,.
\end{equation}
The remaining (even) generators of the supergroup $D(2,1;\alpha)$ can be found by
evaluating mutual anticommutators of the odd generators.

As follows from the action~(\ref{4N-ph}), the ${\rm{SU}}(2)$ spinor
variables are
canonically
self-conjugate due to the presence of second-class constraints for their
momenta.
As a result, non-vanishing canonical Dirac brackets (at equal times) have the
following form
\begin{equation}\label{CDB}
[x, p]_{{}_D}= 1, \qquad [z^i, \bar z_j]_{{}_D}= -i\delta^i_j, \qquad \{\psi^{ii^\prime},
\psi^{kk^\prime}\}_{{}_D}= {\textstyle\frac{i}{2}}\,\epsilon^{i k}\epsilon^{i^\prime
k^\prime} \quad \left(\{\psi^i, \bar\psi_j\}_{{}_D}=
{\textstyle\frac{i}{2}}\,\delta^i_j\right)
\end{equation}
where we introduced the notations
\begin{equation}\label{psi-22}
\psi^{ii^\prime}=(\psi^{i1^\prime},\psi^{i2^\prime})=(\psi^i,
\bar\psi^i),\qquad
(\overline{\psi^{ii^\prime}})=\psi_{ii^\prime}=\epsilon_{i
k}\epsilon_{i^\prime
k^\prime}
\psi^{kk^\prime}\,, \quad (\epsilon_{12} = \epsilon^{21} = 1).
\end{equation}

Using Dirac brackets~(\ref{CDB}), we arrive at the following closed
superalgebra:
\begin{equation}
\{{Q}^{ai^\prime i}, {Q}^{bk^\prime k}\}_{{}_D}= 2i\left(\epsilon^{ik}\epsilon^{i^\prime
k^\prime} T^{ab}+\alpha \epsilon^{ab}\epsilon^{i^\prime k^\prime} J^{ik}-(1+\alpha)
\epsilon^{ab}\epsilon^{ik} I^{i^\prime k^\prime}\right)\,,\label{DB-Q-g}
\end{equation}
\begin{equation}
[T^{ab}, T^{cd}]_{{}_D}= -\epsilon^{ac}T^{bd} -\epsilon^{bd}T^{ac},
\end{equation}
\begin{equation}
[J^{ij}, J^{kl}]_{{}_D}= -\epsilon^{ik}J^{jl} -\epsilon^{jl}J^{ik},\qquad [I^{i^\prime
j^\prime}, I^{k^\prime l^\prime}]_{{}_D}= -\epsilon^{ik}I^{j^\prime l^\prime}
-\epsilon^{j^\prime l^\prime}I^{i^\prime k^\prime},\label{DB-J}
\end{equation}
\begin{equation}
[T^{ab}, {Q}^{ci^\prime i}]_{{}_D}=\epsilon^{c(a}{Q}^{b)i^\prime i} ,\qquad [J^{ij},
{Q}^{ai^\prime k}]_{{}_D}=\epsilon^{k(i}{Q}^{ai^\prime j)},\qquad [J^{i^\prime j^\prime},
{Q}^{ak^\prime i}]_{{}_D}=\epsilon^{k^\prime (i^\prime}{Q}^{aj^\prime ) i}\,. \label{DB-JQ}
\end{equation}
In~(\ref{DB-Q-g})-(\ref{DB-JQ}) we use the notation
\begin{equation}\label{not-Q}
{Q}^{21^\prime i}=-{Q}^{i}\,,\quad {Q}^{22^\prime i}=-\bar{Q}^{i}\,,
\qquad\qquad
{Q}^{11^\prime i}={S}^{i}\,,\quad {Q}^{12^\prime i}=\bar{S}^{i}\,,
\end{equation}
\begin{equation}\label{not-T}
T^{22}=H\,,\quad T^{11}=K\,,\quad T^{12}=-D\,.
\end{equation}
The explicit expressions for the generators are
\begin{eqnarray}\label{H-cl}
H &=&{\textstyle\frac{1}{4}}\,p^2  +\alpha^2\,\frac{(\bar z_k z^{k})^2}{4x^2} - 2\alpha  \,
\frac{\psi^{i}\bar\psi^{k} z_{(i} \bar z_{k)}}{x^2} - (1+2\alpha) \,
\frac{\psi_{i}\psi^{i}\,\bar\psi^{k} \bar\psi_{k}}{2x^2}\,,
\\ \label{K-cl}
K &=&x^2  -t\,x p +
    t^2\, H\,,
\\ \label{D-cl}
D &=&-{\textstyle\frac{1}{2}}\,x p +
    t\, H\,,
\\ \label{T-cl}
J^{ij} &=& i\left[ z^{(i} \bar z^{j)}+ \psi^{ik^\prime}\psi^{j}{}_{k^\prime}\right]=
i\left[ z^{(i} \bar z^{k)}+ 2\psi^{(i}\bar\psi{}^{k)}_{\phantom{k^\prime}}\right]\,,
\\ \label{I-cl}
I^{i^\prime j^\prime} &=& i \psi^{ki^\prime}\psi_{k}{}^{j^\prime}\qquad \left( I^{1^\prime
1^\prime} = -i\psi_k\psi^k\,,\quad I^{2^\prime 2^\prime} = i\bar\psi^k\bar\psi_k\,,\quad
I^{1^\prime 2^\prime} =-i \psi_k\bar\psi^k \right)\,.
\end{eqnarray}
The relations~(\ref{DB-Q-g})-(\ref{DB-JQ}) provide the standard form of the superalgebra
$D(2,1;\alpha)$ (see, e.g.,~\cite{Sorba,BILS,IKLech}). Bosonic generators $T^{ab}=T^{ba}$,
$J^{ik}=J^{ki}$, $I^{i^\prime k^\prime}=I^{k^\prime i^\prime}$ form mutually commuting
$su(1,1)$, $su(2)_R$ and $su(2)_L$ algebras, respectively.\footnote{
It would be of interest to clarify the precise relation of  our
realization of $D(2,1;\alpha)$ derived from the concrete model to the realization found recently
in \cite{HKLN} from a different reasoning.}

It is worth pointing out one important feature of the basic relation
$\{{Q}^{i}, \bar{Q}_{j}\}_{{}_D}= 2iH\delta^i_j$. Although $Q$ and $\bar Q$ contain terms of
the third order in $\psi$ with the coefficients $(1+2\alpha)$,
no quartic fermionic term $\sim(1 +2\alpha)^2$ appears
in the Hamiltonian. This is because of the vanishing Dirac bracket
\begin{equation}\label{0rel-cl}
\{\psi_{k}\psi^{k}\bar\psi^{i} , \bar\psi^{l} \bar\psi_{l}\psi_{j}\}_{{}_D}=0\,.
\end{equation}

The expression~(\ref{H-cl}) coincides with the canonical Hamiltonian
associated with the action~(\ref{4N-ph}).
Owing to the $A$-term in~(\ref{4N-ph}), there is also the first-class
constraint
\begin{equation}\label{D0-con}
D^0 -c\equiv \bar z_k z^{k} -c \approx 0\,,
\end{equation}
which should be imposed on the wave functions in quantum case.

Casimir operators (on classical level) of the $su(1,1)$, $su(2)_R$ and $su(2)_L$ algebras are
\begin{eqnarray}\label{Cas-11}
T^2&\equiv& {\textstyle\frac{1}{2}}\, T^{ab}T_{ab} = HK -D^2 =
{\textstyle\frac{1}{4}} \alpha^2 (z^{k}\bar z_k )^2 -2\alpha\,  z^{(i} \bar
z^{k)}\psi_{(i}\bar\psi_{k)} -{\textstyle\frac{1}{2}}
(1+2\alpha) \psi_{i}\psi^{i}\,\bar\psi^{k} \bar\psi_{k},  \\
\label{Cas-2} J^2&\equiv& {\textstyle\frac{1}{2}}\, J^{ik}J_{ik} =
{\textstyle\frac{1}{4}}\,(z^{k}\bar z_k )^2 - 2 z^{(i} \bar z^{k)}\psi_{(i}\bar\psi_{k)}-
{\textstyle\frac{3}{2}}\,\psi_i\psi^i\, \bar\psi^k\bar\psi_k\,,
\\ \label{Cas-1}
I^2&\equiv& {\textstyle\frac{1}{2}}\, I^{i^\prime k^\prime}I_{i^\prime k^\prime} = I\bar I
-(I_3)^2 = {\textstyle\frac{3}{2}}\,\psi_i\psi^i\, \bar\psi^k\bar\psi_k\,.
\end{eqnarray}
Using these expressions and
\begin{equation}\label{QQ-clCas}
{\textstyle\frac{i}{4}}\, Q^{ai^\prime i}Q_{ai^\prime i} =
{\textstyle\frac{i}{2}}\,(Q^{i}\bar S_{i}- S^{i}\bar Q_{i}) =4\alpha\, z^{(i} \bar
z^{k)}\psi_{(i}\bar\psi_{k)} +2(1+2\alpha) \,\psi_{i}\psi^{i}\,\bar\psi^{k} \bar\psi_{k}\,,
\end{equation}
we obtain that the second-order (classical) Casimir operator of $D(2,1;\alpha)\,$,
\begin{equation}\label{cl-Cas}
C_2=T^2+\alpha J^2- (1+\alpha)I^2 +{\textstyle\frac{i}{4}}\, Q^{ai^\prime i}Q_{ai^\prime i}\,,
\end{equation}
takes the form
\begin{equation}\label{cl-Cas-12}
C_2={\textstyle\frac{1}{4}}\,\alpha(\alpha+1)\,(z^{k}\bar z_k )^2
={\textstyle\frac{1}{4}}\,\alpha(\alpha+1)\,(D^0)^2\,.
\end{equation}

It is important to note that the (iso)spin (angular) variables make significant contributions
to $D(2,1;\alpha)$, $su(1,1)$ and $su(2)_R$  Casimirs (\ref{Cas-11}), (\ref{Cas-2}), (\ref{cl-Cas-12}).
Additional terms in these operators are generated by the second and third terms in the Hamiltonian (\ref{H-cl})
and the first terms in the generators (\ref{T-cl}), all arising from the terms $\propto b$ in the
actions \p{bose} and \p{fermi}.

By inspecting the expressions~(\ref{Cas-11})--(\ref{QQ-clCas}), we observe
that the following quantity $M$ vanishes identically for this particular realization of the $D(2,1;\alpha)$
superalgebra:
\begin{equation}\label{M-cl}
{M}\equiv {T}^2 - \alpha^2 {J}^2-  {\textstyle\frac{1}{3}}\,(1- \alpha^2)\,{I}^2
+{\textstyle\frac{i}{8}}\, (1-\alpha)\, {Q}^{ai^\prime i} {Q}_{ai^\prime i} =0\,.
\end{equation}
Using this identity together with the expression~(\ref{cl-Cas}), we obtain the constraint
\begin{equation}\label{cl-constr}
(\alpha +1)\Big[T^2  - \alpha J^2 -{\textstyle\frac{1}{3}}\,(\alpha -1) I^2\Big]
-(\alpha -1) C_2=0 \,,
\end{equation}
which relates the Casimir of $D(2,1;\alpha)$ to the Casimirs
of the three mutually commuting bosonic subgroups SU(1,1), SU(2)$_L$ and SU(2)$_R$
in our model. Plugging the expression~(\ref{cl-Cas-12}) for the $D(2,1;\alpha)$ Casimir
in this constraint, we find that
\begin{equation}\label{cl-constr-1}
(\alpha +1)\Big[T^2  - \alpha J^2 -{\textstyle\frac{1}{3}}\,(\alpha -1) I^2
-{\textstyle\frac{1}{4}}\,\alpha(\alpha-1)\,(D^0)^2\Big]
=0 \,.
\end{equation}
Using the expressions~(\ref{Cas-11})--(\ref{Cas-1}), we can check that the term
in the square brackets is vanishing, that is the expression
\begin{equation}\label{H-constr-1}
T^2  = \alpha J^2 +{\textstyle\frac{1}{3}}\,(\alpha -1) I^2
+{\textstyle\frac{1}{4}}\,\alpha(\alpha-1)\,(D^0)^2
\end{equation}
is valid for all $\alpha\neq 0\,$, including $\alpha{=}-1\,$.

Note that the Hamiltonian~(\ref{H-cl}) has the standard form of the Hamiltonian of (super)conformal mechanics
\footnote{From $H=\frac{1}{4}(p^2 +\frac{g}{x^2})$ and the expressions~(\ref{K-cl}), (\ref{D-cl})
we obtain $T^2=g/4$.}
\begin{equation}\label{H-cl-st}
H ={\textstyle\frac{1}{4}}\,p^2  +
\frac{T^2}{x^2} \,.
\end{equation}
Using the expression~(\ref{H-constr-1}), we can
represent the Hamiltonian in the convenient equivalent form
\begin{equation}\label{H-cl-a}
H ={\textstyle\frac{1}{4}}\,p^2  +\alpha(\alpha -1)\,\frac{(D^0)^2}{4x^2} + \alpha \,
\frac{J^2}{x^2} + (\alpha -1) \, \frac{I^2}{3x^2}\,.
\end{equation}
The last two terms involve the Casimirs of the groups SU(2)$_R$ and SU(2)$_L$.
The second term contains the quantity $D^0{=} \bar z_k z^{k}$ which is the generator
of some extra U(1) commuting with $D(2,1;\alpha)\,$.

It is worth pointing out that at $\alpha{=}-1$, when $D(2,1;\alpha)$ degenerates into
${\rm SU(1,1|2)}{\subset\!\!\!\!\!\!\times}{\rm SU(2)}_L$, the SU(2)$_L$ Casimir $I^2$ drops out from
the expression \p{cl-Cas} for the Casimir $C_2$, as it should be.
However, since in the model under consideration this SU(2)$_L$ is realized only on fermions,
the Casimir $I^2$ reappears in the subsequent formulas
from the term ${\textstyle\frac{i}{4}}\, Q^{ai^\prime i}Q_{ai^\prime i}$. Hence,
even for a fixed $D(2,1;\alpha)$ Casimir~(\ref{cl-Cas}),
the term ${\textstyle\frac{i}{4}}\, Q^{ai^\prime i}Q_{ai^\prime i}$ makes a contribution $\sim I^2$ to
the ${\rm SU}(1,1)$ Casimir~(\ref{H-constr-1}).
As a result, the term  with the ${\rm SU(2)}_{L}$ Casimir $I^2$
is retained in~(\ref{H-cl-a}) even at $\alpha{=}-1\,$.
Incidentally, the simplest form of the Hamiltonian is achieved at $\alpha{=}1\,$.

In the next section we shall construct a quantum realization of the $D(2,1;\alpha)$
superalgebra.

\setcounter{equation}0
\section{$D(2,1;\alpha)$ quantum mechanics}

\subsection{Operator realization of $D(2,1;\alpha)$ superalgebra}

Quantum operators of physical coordinates and momenta satisfy the quantum
brackets,
obtained in the standard way from~(\ref{CDB})
\begin{equation}\label{cB}
[X, P] = i\,, \qquad [Z^i, \bar Z_j] = \delta^i_j \,, \qquad \{\Psi^i,
\bar\Psi_j\}=
-{\textstyle\frac{1}{2}}\,\delta^i_j \,.
\end{equation}

Quantum supertranslation and superconformal boost generators are defined by the classical
expressions (\ref{Q-cl}), (\ref{Qb-cl}), (\ref{S-cl}). We take Weyl ordering of
the fermionic quantities in the last terms of (\ref{Q-cl}) and (\ref{Qb-cl}):
\begin{equation}\label{Q-qu}
\mathbf{Q}^i =P \Psi^i+ 2i\alpha\frac{Z^{(i} \bar Z^{k)}\Psi_k}{X}+
i(1+2\alpha)\frac{\langle\Psi_{k} \Psi^{k}\bar\Psi^i\rangle}{X}\, ,
\end{equation}
\begin{equation}\label{Qb-qu}
\bar\mathbf{Q}_i=P \bar\Psi_i- 2i\alpha\frac{Z_{(i} \bar Z_{k)}\bar\Psi^k}{X} +
i(1+2\alpha)\frac{\langle\bar\Psi^{k} \bar\Psi_{k}\Psi_i\rangle}{X}\,,
\end{equation}
\begin{equation}\label{S-qu}
\mathbf{S}^i =-2\,X \Psi^i + t\,\mathbf{Q}^i,\qquad \bar\mathbf{S}_i=-2\,X
\bar\Psi_i+ t\,\bar\mathbf{Q}_i\,.
\end{equation}
The symbol $\langle...\rangle$ denotes Weyl ordering. Note that
$$
\langle\Psi_{k} \Psi^{k}\bar\Psi^i\rangle =\Psi_{k}
\Psi^{k}\bar\Psi^i+{\textstyle\frac12}\,\Psi^i\,,\qquad \langle\bar\Psi^{k}
\bar\Psi_{k}\Psi_i\rangle= \bar\Psi^{k} \bar\Psi_{k}\Psi_i +{\textstyle\frac12}\,
\bar\Psi_i
$$
and $\bar\mathbf{Q}_i=-\left(\mathbf{Q}^i\right)^+$,
$\bar\mathbf{S}_i=-\left(\mathbf{S}^i\right)^+$.

Evaluating the anticommutators of the odd generators (\ref{Q-qu}), (\ref{S-qu}), one
determines uniquely the full set of quantum generators of superconformal algebra
$D(2,1;\alpha)$. We obtain \footnote{It is worth making here an important clarifying remark which refers as well to our
previous paper \cite{FIL-09}. In ~(\ref{cB}) and below we assign to quantum operators the following
Hermitian conjugation properties
\begin{equation}\label{Her}
X^+=X,\quad P^+ = P\,, \qquad \bar Z_i = -\left(Z^i\right)^+  \,, \qquad \bar\Psi_i=
-\left(\Psi^i\right)^+\,,
\end{equation}
whereas for classical quantities we still have $\bar z_i = \overline{\left(z^i\right)}$,
$\bar\psi_i= \overline{\left(\psi^i\right)}$. This change of conventions in the quantum case is necessary
for ensuring the standard  Clifford algebra for quantum fermionic operators and standard quantum supersymmetry
algebra with the positive-definte right-hand side of the basic anticommutator (see the
comments after (\ref{qB-Q-g})-(\ref{not-Tq})). As we show in Appendix B, the standard conjugation conventions
can be restored by performing the time reversal $t\rightarrow -t$ in the initial model, thus bringing the opposite (standard)
sign to kinetic terms of all involved $d{=}1$ spinor fields.}
\begin{eqnarray}
\mathbf{H} &=&{\textstyle\frac{1}{4}}\,P^2  +\alpha^2\frac{(\bar Z_k Z^{k})^2+2\bar Z_k
Z^{k}}{4X^2} - 2\alpha    \frac{Z^{(i} \bar Z^{k)} \Psi_{(i}\bar\Psi_{k)}}{X^2} \label{H-qu}\\
&& -\, (1+2\alpha)
\frac{\langle\Psi_{i}\Psi^{i}\,\bar\Psi^{k} \bar\Psi_{k}\rangle}{2X^2}+
  \frac{(1+2\alpha)^2}{16X^2}\,,\nonumber
\\ \label{K-qu}
\mathbf{K} &=&X^2  - t\,{\textstyle\frac{1}{2}}\,\{X, P\} +
    t^2\, \mathbf{H}\,,
\\ \label{D-qu}
\mathbf{D} &=&-{\textstyle\frac{1}{4}}\,\{X, P\} +
    t\, \mathbf{H}\,,
\\ \label{T-qu}
\mathbf{J}^{ik} &=& i\left[ Z^{(i} \bar Z^{k)}+
2\Psi^{(i}\bar\Psi^{k)}\right]\,,
\\ \label{I-qu}
\mathbf{I}^{1^\prime 1^\prime} &=& -i\Psi_k\Psi^k\,,\qquad
\mathbf{I}^{2^\prime
2^\prime} =
i\bar\Psi^k\bar\Psi_k\,,\qquad \mathbf{I}^{1^\prime 2^\prime}
=-{\textstyle\frac{i}{2}}\,
[\Psi_k,\bar\Psi^k]\,.
\end{eqnarray}
Note that
$$
\langle\Psi_{i}\Psi^{i}\,\bar\Psi^{k} \bar\Psi_{k}\rangle
={\textstyle\frac12}\,\left\{\Psi_{i}\Psi^{i},\bar\Psi^{k} \bar\Psi_{k}\right\}
-{\textstyle\frac{1}{4}} =\Psi_{i}\Psi^{i}\,\bar\Psi^{k} \bar\Psi_{k} -
\Psi_{i}\bar\Psi^{i}  +{\textstyle\frac{1}{4}} \,,
$$
$$
\Psi^{i} \langle\Psi_{l}\Psi^{l}\,\bar\Psi^{k} \bar\Psi_{k}\rangle
=- \langle\Psi_{l}\Psi^{l}\,\bar\Psi^{k} \bar\Psi_{k}\rangle \Psi^{i}=
{\textstyle\frac12}\,\langle\Psi_{l}\Psi^{l} \bar\Psi^{i} \rangle \,,
$$
$$
\bar\Psi_{i} \langle\Psi_{l}\Psi^{l}\,\bar\Psi^{k} \bar\Psi_{k}\rangle
=- \langle\Psi_{l}\Psi^{l}\,\bar\Psi^{k} \bar\Psi_{k}\rangle \bar\Psi_{i}=
{\textstyle\frac12}\,\langle\bar\Psi^{k}\bar\Psi_{k} \Psi_{i} \rangle \,.
$$
It can be directly checked that the generators~(\ref{Q-qu})--(\ref{I-qu}) indeed form the
$D(2,1;\alpha)$ superalgebra which is obtained from the DB
superalgebra~(\ref{DB-Q-g})-(\ref{DB-JQ}) in the standard fashion (changing altogether DB by (anti)commutators
and multiplying the right-hand sides by $i$):
\begin{equation}
\{\mathbf{Q}^{ai^\prime i}, \mathbf{Q}^{bk^\prime k}\}=
-2\left(\epsilon^{ik}\epsilon^{i^\prime k^\prime} \mathbf{T}^{ab}+\alpha
\epsilon^{ab}\epsilon^{i^\prime k^\prime} \mathbf{J}^{ik}-(1+\alpha)
\epsilon^{ab}\epsilon^{ik} \mathbf{I}^{i^\prime k^\prime}\right)\,,\label{qB-Q-g}
\end{equation}
\begin{equation}
[\mathbf{T}^{ab}, \mathbf{T}^{cd}]= -i\left(\epsilon^{ac}\mathbf{T}^{bd}
+\epsilon^{bd}\mathbf{T}^{ac}\right)\,,
\end{equation}
\begin{equation}
[\mathbf{J}^{ij}, \mathbf{J}^{kl}]= -i\left(\epsilon^{ik}\mathbf{J}^{jl}
+\epsilon^{jl}\mathbf{J}^{ik}\right)\,,\qquad [\mathbf{I}^{i^\prime j^\prime},
\mathbf{I}^{k^\prime l^\prime}]= -i\big(\epsilon^{ik}\mathbf{I}^{j^\prime l^\prime}
+\epsilon^{j^\prime l^\prime}\mathbf{I}^{i^\prime k^\prime}\big)\,,\label{qB-J}
\end{equation}
\begin{equation}
[\mathbf{T}^{ab}, \mathbf{Q}^{ci^\prime i}]=i\epsilon^{c(a}\mathbf{Q}^{b)i^\prime i}
,\qquad [\mathbf{J}^{ij}, \mathbf{Q}^{ai^\prime k}]=i\epsilon^{k(i}\mathbf{Q}^{ai^\prime
j)},\qquad [\mathbf{J}^{i^\prime j^\prime}, \mathbf{Q}^{ak^\prime i}]=i\epsilon^{k^\prime
(i^\prime}\mathbf{Q}^{aj^\prime ) i}\,. \label{qB-JQ}
\end{equation}
As in~(\ref{DB-Q-g})-(\ref{DB-JQ}), in~(\ref{qB-Q-g})-(\ref{qB-JQ}) we use the notation
\begin{equation}\label{not-Qq}
\mathbf{Q}^{21^\prime i}=-\mathbf{Q}^{i}\,,\quad \mathbf{Q}^{22^\prime
i}=-\bar\mathbf{Q}^{i}\,, \qquad\qquad \mathbf{Q}^{11^\prime i}=\mathbf{S}^{i}\,,\quad
\mathbf{Q}^{12^\prime i}=\bar\mathbf{S}^{i}\,,
\end{equation}
\begin{equation}\label{not-Tq}
\mathbf{T}^{22}=\mathbf{H}\,,\quad \mathbf{T}^{11}=\mathbf{K}\,,\quad
\mathbf{T}^{12}=-\mathbf{D}\,.
\end{equation}
Note that due to~(\ref{Her}) we have
\begin{equation}\label{Her-Q}
\left(\mathbf{Q}^{ai^\prime i}\right)^+ = -\epsilon_{ik}\epsilon_{i^\prime
k^\prime}\mathbf{Q}^{ak^\prime k}
\end{equation}
and, as a result, the basic anticommutator has the standard form
$\{\mathbf{Q},{\mathbf{Q}}^+\}=\mathbf{H}\,$.

In the quantum case, the classical relation~(\ref{0rel-cl}) is replaced by
\begin{equation}\label{0rel-qu}
\{\langle\Psi_{k}\Psi^{k}\bar\Psi^{i}\rangle , \langle\bar\Psi^{l} \bar\Psi_{l}\Psi_{j}\rangle\}={\textstyle\frac{1}{8}}\,\delta^i_j\,.
\end{equation}
and, due to~(\ref{0rel-qu}), the term $\frac{(1+2\alpha)^2}{16X^2}$ appears in the quantum Hamiltonian~(\ref{H-qu}).
This term is necessary also for preserving the basic supersymmetry relations
$[\mathbf{H},\mathbf{Q}]=[\mathbf{H},\bar{\mathbf{Q}}]= 0$. The appearance of such a ``conformal'' term when quantizing ${\cal N}{=}4$
superconformal systems was earlier observed in~\cite{GLP}.

The quantization of the pure bosonic limit~(\ref{bose}) of the classical system~(\ref{4N-ph})
does not lead to appearance of the additional term $\frac{(1+2\alpha)^2}{16X^2}$ in the corresponding quantum Hamiltonian
which is thus a sum of only first two terms in~(\ref{H-qu}).
Using the same procedure as in~\cite{FIL-09} this Hamiltonian can be represented in the form
\begin{equation}\label{H-qu-bose}
H =  \frac{1}{4} \left[ P^2 + 4\alpha^2\frac{Y_a Y_a}{X^2} \right],
\end{equation}
where
\begin{equation}\label{qu-y}
Y_a= {\textstyle\frac{1}{2}}\,\bar Z_i(\sigma_a)^i{}_j Z^j
\end{equation}
and $\sigma_a$, $a=1,2,3$ are Pauli matrices. The quantities $Y_a$,
obtained via the first Hopf map from the ${\rm SU}(2)$ spinors $Z^i, \bar Z_i$, generate ${\rm SU(2)}_R$
transformations in the bosonic sector of the model
(the second ${\rm SU(2)}_L$ R-symmetry group of $D(2,1;\alpha)$ acts in the fermionic sector only).
The operator $Y_a Y_a$ in the second term of~(\ref{H-qu-bose}) is the Casimir operator of the group ${\rm SU(2)}_R$
for its realization in the bosonic sector.
Due to the constraint~(\ref{D0-con}) (for definiteness, we adopt $\bar Z_k
Z^{k}$--ordering in it; see also~(\ref{op-D0}) and (\ref{q-con})), this Casimir takes
the definite value
$\frac{c}{2} \left(\frac{c}{2}+1\right)$.
Thus, in the pure bosonic limit our model describes a conformal particle with the quantum potential
$\alpha^2 \frac{c}{2} \left(\frac{c}{2}+1\right)/X^2$ which possesses the fixed ${\rm SU(2)}_R$
spin $\frac{c}{2}$. In the entire supersymmetric model, with all fermions taken into account,
the generators of ${\rm SU(2)}_R$ contain additional fermionic parts (see~(\ref{T-qu})) and the corresponding
full ${\rm SU(2)}_R$ Casimir operator proves not to be fixed. A thorough consideration of the pure bosonic case
of the $\alpha=-1/2$ model can be found in our paper~\cite{FIL-09}.

 The second-order Casimir operator of the whole supergroup $D(2,1;\alpha)$ is given by the following
expression~\cite{Je}
\begin{equation}\label{qu-Cas}
\mathbf{C}_2=\mathbf{T}^2 +\alpha\, \mathbf{J}^2- (1+\alpha)\,\mathbf{I}^2 +
{\textstyle\frac{i}{4}}\, \mathbf{Q}^{ai^\prime i}\mathbf{Q}_{ai^\prime i}\,.
\end{equation}
Using the relations
\begin{eqnarray}\label{q-Cas-3}
\mathbf{T}^2&\equiv& {\textstyle\frac{1}{2}}\,
\mathbf{T}^{ab}\mathbf{T}_{ab}={\textstyle\frac{1}{2}}\,\{\mathbf{H},\mathbf{K}\}
-\mathbf{D}^2 = {\textstyle\frac{1}{4}}\,\alpha^2 \left[(\bar Z_k Z^{k})^2+2\bar Z_k
Z^{k}\right] -2\alpha Z^{(i} \bar Z^{k)}\Psi_{(i}\bar\Psi_{k)}  \\
&& \qquad\qquad\qquad\qquad\qquad\qquad\quad - {\textstyle\frac{1}{2}}\,(1+2\alpha)
\langle\Psi_{i}\Psi^{i}\,\bar\Psi^{k} \bar\Psi_{k}\rangle +
  {\textstyle\frac{1}{16}}\,(1+2\alpha)^2 -{\textstyle\frac{3}{16}}\,, \nonumber
\\
\label{q-Cas-2}
\mathbf{J}^2&\equiv& {\textstyle\frac{1}{2}}\,
\mathbf{J}^{ik}\mathbf{J}_{ik} =
{\textstyle\frac{1}{4}}\, \left[(\bar Z_k Z^{k})^2+2\bar Z_k Z^{k}\right] -
{\textstyle\frac{3}{2}}\,\left(\Psi_i\Psi^i\, \bar\Psi^k\bar\Psi_k -
\Psi_i \bar\Psi^i
\right) - 2 Z^{(i} \bar Z^{k)}\Psi_{(i}\bar\Psi_{k)},
\\
\label{q-Cas-1}
\mathbf{I}^2&\equiv& {\textstyle\frac{1}{2}}\, \mathbf{I}^{i^\prime k^\prime
}\mathbf{I}_{i^\prime k^\prime }
={\textstyle\frac{1}{2}}\,\{\bar\mathbf{I},\mathbf{I}\} -(\mathbf{I}_3)^2
= {\textstyle\frac{3}{2}}\,\left(\Psi_i\Psi^i\, \bar\Psi^k\bar\Psi_k - \Psi_i
\bar\Psi^i
\right)+ {\textstyle\frac{3}{4}}
\end{eqnarray}
together with
\begin{eqnarray}\label{QQ-quCas}
{\textstyle\frac{i}{4}}\, \mathbf{Q}^{ai^\prime i}\mathbf{Q}_{ai^\prime i} &=&
{\textstyle\frac{i}{4}}\, [\mathbf{Q}^{i},\bar \mathbf{S}_{i}] + {\textstyle\frac{i}{4}}\,
[\bar \mathbf{Q}_{i}, \mathbf{S}^{i}] \\
&=& 4\alpha Z^{(i} \bar Z^{k)}\Psi_{(i}\bar\Psi_{k)} + 2(1+2\alpha)\left(\Psi_i\Psi^i\,
\bar\Psi^k\bar\Psi_k - \Psi_i \bar\Psi^i \right)
  +(1+\alpha)\,,\nonumber
\end{eqnarray}
we finally cast $\mathbf{C}_2$ in the form
\begin{equation}\label{qu-Cas-12}
\mathbf{C}_2={\textstyle\frac{1}{4}}\, \alpha(1+\alpha)\Big[(\bar Z_k Z^{k})^2+2\bar Z_k
Z^{k} +1 \Big] \,.
\end{equation}

\subsection{Invariant spaces in the enveloping
algebra of $D(2,1;\alpha)$}

An important property is that the enveloping algebra of $D(2,1;\alpha)$ superalgebra
has several subspaces which are closed under the action of $D(2,1;\alpha)$.
The presence of such subspaces provides an explanation why some bilinear combinations
of the $D(2,1;\alpha)$ generators in the considered realization identically vanish without conflict with
the $D(2,1;\alpha)$ covariance. This phenomenon is encountered already at the
classical level (see~(\ref{M-cl})). As we shall see, the realization of the $D(2,1;\alpha)$ generators in the considered model is such
that the operators forming one of the invariant subspaces just mentioned are vanishing. As a result,
the physical states form a module of such a restricted representation of $D(2,1;\alpha)$.

One invariant subspace is formed by the bilinear combinations
\begin{eqnarray}\label{M}
\mathbf{M}&\equiv&  \mathbf{T}^2 - \alpha^2\,
 \mathbf{J}^2-  {\textstyle\frac{1}{3}}\,(1-\alpha^2)\, \mathbf{I}^2 + {\textstyle\frac{i}{8}}\,(1-\alpha)\, \mathbf{Q}^{ai^\prime
i}\mathbf{Q}_{ai^\prime i} \,,
\\
\label{M-3}
\mathbf{M}^{ai^\prime i}&\equiv&  {\textstyle\frac{i}{4}}\, \left(
\{ \mathbf{T}^{a}_b , \mathbf{Q}^{bi^\prime i} \} -
\alpha\,
\{ \mathbf{J}^{i}_j, \mathbf{Q}^{ai^\prime j}\} +
{\textstyle\frac{1}{3}}\,(1-\alpha)\,
\{ \mathbf{I}^{i^\prime}_{j^\prime}, \mathbf{Q}^{aj^\prime i} \} \right)\,,
\\
\label{M-4}
\mathbf{M}^{ik\!,\,i^\prime k^\prime}&\equiv&  \alpha\, \{ \mathbf{J}^{ik},
\mathbf{I}^{i^\prime k^\prime}\}
- {\textstyle\frac{i}{2}} \, \mathbf{Q}^{b{}^{\scriptstyle (}i^\prime (i}\mathbf{Q}_{b}{}^{k^\prime
{}^{\scriptstyle )}
k )} \,,
\\
\label{M-4a}
\mathbf{M}^{ac\!,\,i^\prime k^\prime}&\equiv&  \{ \mathbf{T}^{ac},
\mathbf{I}^{i^\prime k^\prime}\}
- {\textstyle\frac{i}{2}} \,\mathbf{Q}^{{}^{\scriptstyle (}a (i^\prime j}\mathbf{Q}^{c
{}^{\scriptstyle )}k^\prime )}{}_{j} \,,
\\
\label{M-5}
\mathbf{M}^{ai\!,\,i^\prime j^\prime k^\prime}&\equiv&  i \{ \mathbf{I}^{(i^\prime j^\prime },
\mathbf{Q}^{ak^\prime ) k }\} \,,
\\
\label{M-6}
\mathbf{M}^{i^\prime j^\prime k^\prime l^\prime}&\equiv&   \{ \mathbf{I}^{(i^\prime j^\prime },
\mathbf{I}^{k^\prime l^\prime )}\} \,.
\end{eqnarray}
On this set a linear finite-dimensional representation of $D(2,1;\alpha)$ is realized
\begin{eqnarray}\label{Ma1}
[\mathbf{M}, \mathbf{Q}^{ai^\prime i}]&=& (1+\alpha)\,\mathbf{M}^{ai^\prime i} \,,
\\
\label{Ma2}
\{\mathbf{M}^{ai^\prime i}, \mathbf{Q}^{ck^\prime
k}\} &=&  -i\,
\epsilon^{ac}
\epsilon^{i^\prime k^\prime}\epsilon^{ik}\mathbf{M}
+ {\textstyle\frac{i}{3}}\,(2+\alpha)\,
\epsilon^{ac}\mathbf{M}^{ik\!,\,i^\prime k^\prime} - {\textstyle\frac{i}{3}}\,(1+2\alpha)\,
\epsilon^{ik}\mathbf{M}^{ac\!,\,i^\prime k^\prime},
\\
\label{Ma3}
[\mathbf{M}^{ik\!,\,i^\prime k^\prime}, \mathbf{Q}^{bj^\prime
j}] &=&  4\epsilon^{j(i}
\epsilon^{j^\prime{}^{\scriptstyle
(}i^\prime}\mathbf{M}^{bk^\prime{}^{\scriptstyle
)} k)} +(1+2\alpha)\epsilon^{j(i}
\mathbf{M}^{b k),\,i^\prime j^\prime k^\prime}\,,
\\
\label{Ma4}
[\mathbf{M}^{ac\!,\,i^\prime k^\prime}, \mathbf{Q}^{bj^\prime
j}] &=&  -4\epsilon^{b(a}
\epsilon^{j^\prime{}^{\scriptstyle
(}i^\prime}\mathbf{M}^{c) k^\prime{}^{\scriptstyle)}j} +(2+\alpha)\epsilon^{b(a}
\mathbf{M}^{c) j,\,i^\prime j^\prime k^\prime} \,,
\\
\label{Ma5}
\{\mathbf{M}^{ai\!,\,i^\prime j^\prime k^\prime}, \mathbf{Q}^{bl^\prime
l}\} \!\!&=&  -2i\epsilon^{ba} \epsilon^{l^\prime(i^\prime}\mathbf{M}^{il, j^\prime k^\prime)} -2i\epsilon^{li}
\epsilon^{l^\prime(i^\prime}\mathbf{M}^{ab, j^\prime k^\prime)} +2i(1+\alpha)
\epsilon^{ba} \epsilon^{li}\mathbf{M}^{i^\prime j^\prime k^\prime l^\prime} \,,
\\
\label{Ma6}
[\mathbf{M}^{i^\prime j^\prime k^\prime l^\prime}, \mathbf{Q}^{bn^\prime n}] &=&
\epsilon^{n^\prime(i^\prime}
\mathbf{M}^{bn,\,i^\prime j^\prime k^\prime)} \,.
\end{eqnarray}

The second invariant subspace is formed by the quantities
\begin{eqnarray}\label{N}
\mathbf{N}&\equiv&  \mathbf{T}^2 + {\textstyle\frac{1}{3}}\,\alpha(2+\alpha)\,
 \mathbf{J}^2-  (1+\alpha)^2\, \mathbf{I}^2 + {\textstyle\frac{i}{8}}\,(2+\alpha)\, \mathbf{Q}^{ai^\prime
i}\mathbf{Q}_{ai^\prime i} \,,
\\
\label{N-3}
\mathbf{N}^{ai^\prime i}&\equiv&  {\textstyle\frac{i}{4}}\, \left(
\{ \mathbf{T}^{a}_b , \mathbf{Q}^{bi^\prime i} \} +
{\textstyle\frac{1}{3}}\,(2+\alpha)\,
\{ \mathbf{J}^{i}_j, \mathbf{Q}^{ai^\prime j}\} +
(1+\alpha)\,
\{ \mathbf{I}^{i^\prime}_{j^\prime}, \mathbf{Q}^{aj^\prime i} \} \right)\,,
\\
\label{N-4}
\mathbf{N}^{i^\prime k^\prime\!,\,i k}&\equiv&  -(1+\alpha)\, \{ \mathbf{J}^{ik},
\mathbf{I}^{i^\prime k^\prime}\}
- {\textstyle\frac{i}{2}} \, \mathbf{Q}^{b{}^{\scriptstyle (}i^\prime (i}\mathbf{Q}_{b}{}^{k^\prime
{}^{\scriptstyle )}
k )} \,,
\\
\label{N-4a}
\mathbf{N}^{ac\!,\,i k}&\equiv&  \{ \mathbf{T}^{ac},
\mathbf{J}^{i k}\}
- {\textstyle\frac{i}{2}} \,\mathbf{Q}^{{}^{\scriptstyle (}a j^\prime (i }\mathbf{Q}^{c
{}^{\scriptstyle )}}{}_{j^\prime}{}^{k )} \,,
\\
\label{N-5}
\mathbf{N}^{ai^\prime\!,\,i j k}&\equiv&  i \{ \mathbf{J}^{(i j },
\mathbf{Q}^{ai^\prime k) }\} \,,
\\
\label{N-6}
\mathbf{N}^{i j k l}&\equiv&  \{ \mathbf{J}^{(i j },
\mathbf{J}^{kl)}\}\,.
\end{eqnarray}
They can also be shown to constitute a basis of a linear finite-dimensional representation of $D(2,1;\alpha)\,$.

At last, the third invariant subspace is formed by the bilinear operators
\begin{eqnarray}\label{L}
\mathbf{L}&\equiv&  {\textstyle\frac{1}{3}}\,(1+2\alpha)\,\mathbf{T}^2 + \alpha^2\,
 \mathbf{J}^2-  (1+\alpha)^2\, \mathbf{I}^2 + {\textstyle\frac{i}{8}}\,(1+2\alpha)\, \mathbf{Q}^{ai^\prime
i}\mathbf{Q}_{ai^\prime i} \,,
\\
\label{L-3}
\mathbf{L}^{ai^\prime i}&\equiv&  {\textstyle\frac{i}{4}}\, \left(
{\textstyle\frac{1}{3}}\,(1+2\alpha)\, \{ \mathbf{T}^{a}_b , \mathbf{Q}^{bi^\prime i} \} +
\alpha\,
\{ \mathbf{J}^{i}_j, \mathbf{Q}^{ai^\prime j}\} +
(1+\alpha)\,
\{ \mathbf{I}^{i^\prime}_{j^\prime}, \mathbf{Q}^{aj^\prime i} \} \right)\,,
\\
\label{L-4}
\mathbf{L}^{i^\prime k^\prime\!,\,ac}&\equiv&  -(1+\alpha)\, \{
\mathbf{I}^{i^\prime k^\prime},\mathbf{T}^{ac},\}
- {\textstyle\frac{i}{2}} \, \mathbf{Q}^{{}^{\scriptstyle (}a(i^\prime j}\mathbf{Q}^{c
{}^{\scriptstyle )}k^\prime )}{}_{j} \,,
\\
\label{L-4a}
\mathbf{L}^{ik\!,\,ac}&\equiv&  \alpha\{ \mathbf{J}^{i k}, \mathbf{T}^{ac}\}
- {\textstyle\frac{i}{2}} \,\mathbf{Q}^{{}^{\scriptstyle (}a j^\prime (i }\mathbf{Q}^{c
{}^{\scriptstyle )}}{}_{j^\prime}{}^{k )} \,,
\\
\label{L-5}
\mathbf{L}^{ii^\prime\!,\,abc}&\equiv&  i \{ \mathbf{T}^{(ab },
\mathbf{Q}^{c)i^\prime i }\} \,,
\\
\label{L-6}
\mathbf{L}^{abcd}&\equiv&  \{ \mathbf{T}^{(ab },
\mathbf{T}^{cd)}\}\,.
\end{eqnarray}
As for two previous invariant subspaces, these operators  are closed under the action of $D(2,1;\alpha)\,$.

These three invariant subspaces in the enveloping algebra have the following properties.

First, these subspaces and one-dimensional space formed by the Casimir operator (\ref{qu-Cas})
exhaust all possible invariant subspaces in the enveloping algebra, such that they are bilinear in the $D(2,1;\alpha)\,$
generators and involve singlets of all three bosonic subgroup ${\rm SL}(2,R)$, ${\rm SU(2)}_R$ and ${\rm SU(2)}_L$.

Second, these subspaces are related to each other via some discrete transformations.

Namely, the subspaces
(\ref{M})-(\ref{M-6}) and (\ref{N})-(\ref{N-6}) are {\it dual} to each other. That is,
the discrete transformation
\begin{equation}\label{dual-tr}
\alpha\quad \leftrightarrow \quad -(1+\alpha) \,,
\qquad\qquad
\mathbf{J}^{ik}\quad \leftrightarrow \quad
\mathbf{I}^{i^\prime k^\prime}\,,
\end{equation}
which is an automorphism of the $D(2,1;\alpha)$ algebra~(\ref{qB-Q-g})-(\ref{qB-JQ}), takes the space (\ref{M})-(\ref{M-6})
into the space (\ref{N})-(\ref{N-6}) and vice versa.
The subspace  (\ref{L})-(\ref{L-6})
is a fixed point of the mapping (\ref{dual-tr}).

The subspace (\ref{L})-(\ref{L-6}) is related to the subspaces
(\ref{M})-(\ref{M-6}) and (\ref{N})-(\ref{N-6}) via similar discrete transformations.
E.g., under the transformation
\begin{equation}\label{pseudo-dual-tr}
\alpha\quad \rightarrow \quad \alpha^{-1} \,,
\qquad\quad
\mathbf{T}^{ab}\quad \leftrightarrow \quad
\mathbf{J}^{ik}\,,\qquad\quad \mathbf{Q}^{ai^\prime i}\quad \rightarrow \quad
\alpha^{-1/2}\mathbf{Q}^{ai^\prime i}
\end{equation}
the space (\ref{N})-(\ref{N-6}) goes over into the space (\ref{L})-(\ref{L-6}).
Note, however,  that the change (\ref{pseudo-dual-tr}) (and its analog taking (\ref{M})-(\ref{M-6}) into (\ref{L})-(\ref{L-6}))
is ill defined for the real form of the superalgebra  $D(2,1;\alpha)$ since it takes the $sl(2, R)$ generators into
the $su(2)$ ones.
These transformations present a true automorphism of the complexified $D(2,1;\alpha)$ algebra.

In the case of $\alpha=-1/2$ (when $1+2\alpha=0$)
the subspaces (\ref{M})-(\ref{M-6}) and (\ref{N})-(\ref{N-6}) coincide. Moreover,
the subspace formed by
\begin{equation}\label{sp-12}
\mathbf{M} \,,
\qquad
\mathbf{M}^{ai^\prime i} \,,
\qquad
\mathbf{M}^{ik\!,\,i^\prime k^\prime}
\end{equation}
(or $\mathbf{N}$, $\mathbf{N}^{ai^\prime i}$,
$\mathbf{N}^{i^\prime k^\prime\!,\,i k}$) is invariant under the $D(2,1;\alpha =-1/2)\,$.
Just this subspace was exploited in~\cite{FIL-09}.

In the case of $\alpha=-1$ (when $D(2,1;\alpha){=}{\rm SU}(1,1|2)\!\!\subset\!\!\!\!\!\!\times{\rm SU(2)}_L$)
the operator (\ref{M}) coincides with the Casimir (\ref{qu-Cas}),
\begin{equation}\label{eq-1}
\mathbf{C_2}=\mathbf{M} =\mathbf{T}^2 -
 \mathbf{J}^2 + {\textstyle\frac{i}{4}}\, \mathbf{Q}^{ai^\prime
i}\mathbf{Q}_{ai^\prime i}\,.
\end{equation}
Thus in this special case the appropriate invariant subspaces degenerate into the singlets of
the superconformal group ${\rm SU}(1,1|2)\!\!\subset\!\!\!\!\!\!\times{\rm SU}(2)\,$\footnote{
Although our mechanical system is ill defined at $\alpha{=}0$, the $D(2,1;\alpha)$ algebra \p{qB-Q-g} - \p{qB-JQ}
as it stands still admits such a choice,
and it gives rise  to the superalgebra $D(2,1;\alpha{=}0){=}{\rm SU}(1,1|2)\!\!\subset\!\!\!\!\!\!\times{\rm SU(2)}_R\,$.
In this case the operator (\ref{N}) coincides with the Casimir (\ref{qu-Cas}),
$\mathbf{C_2}=\mathbf{N} =\mathbf{T}^2 -
\mathbf{I}^2 + {\textstyle\frac{i}{4}}\, \mathbf{Q}^{ai^\prime
i}\mathbf{Q}_{ai^\prime i}\,$.}.

Actually, in the case of generic $\alpha$, for the particular representation of generators given by eqs.~(\ref{q-Cas-3})-(\ref{q-Cas-1})
all quantities  (\ref{M})-(\ref{M-6}) identically vanish:
\begin{equation}\label{id-alg}
\mathbf{M}=0 \,,
\quad
\mathbf{M}^{ai^\prime i}=0 \,,
\quad
\mathbf{M}^{ik\!,\,i^\prime k^\prime}=0\,,
\quad
\mathbf{M}^{ac\!,\,i^\prime k^\prime}=0\,,
\quad
\mathbf{M}^{ai\!,\,i^\prime j^\prime k^\prime}=0\,,
\quad
\mathbf{M}^{i^\prime j^\prime k^\prime l^\prime}=0\,.
\end{equation}
As a consequence of these identities, there arises the relation
\begin{equation}\label{id-Cas}
(1+\alpha)\mathbf{T}^2 -\alpha(1+\alpha) \mathbf{J}^2+
{\textstyle\frac{1}{3}}\,(1-\alpha^2)\mathbf{I}^2 =
-(1-\alpha)\mathbf{C}_2\,.
\end{equation}
In the case of $\alpha= -1$ the constraint (\ref{id-Cas}) leads to the condition $\mathbf{C}_2=0$ that agrees with
eqs. (\ref{eq-1}) and \p{id-alg}, as well as with (\ref{qu-Cas-12}).

Using the expression (\ref{qu-Cas-12})
for the Casimir in r.h.s. of (\ref{id-Cas}) we can represent the relation (\ref{id-Cas}) in the form
\begin{equation}\label{id-Cas-1}
\mathbf{T}^2 -\alpha\, \mathbf{J}^2+
{\textstyle\frac{1}{3}}\,(1-\alpha)\mathbf{I}^2 = -\alpha(1-\alpha)
\Big[{\textstyle\frac{1}{2}}\,D^0({\textstyle\frac{1}{2}}\,D^0 +1) +{\textstyle\frac{1}{4}}\, \Big] \,,
\end{equation}
which is valid for any value of $\alpha$. Thus, for an irreducible representation of $D(2,1;\alpha)$
with the fixed $\mathbf{C}_2$ (see
(\ref{qu-Cas-ev}) below), the values of the Casimir operators $\mathbf{T}^2$,
$\mathbf{J}^2$, $\mathbf{I}^2$ of the three bosonic subgroups $sl(2,R)$, $su(2)_R$,
$su(2)_L$ prove to be always related according to \p{id-Cas-1}.

The operator
\begin{equation}\label{op-D0}
D^0=\bar Z_k Z^{k}\,,
\end{equation}
entering the right-hand side of  (\ref{id-Cas-1})
commutes with all generators of the superalgebra $D(2,1;\alpha)\,$ (as in the classical case).

\subsection{Quantum spectrum}

The Hamiltonian (\ref{H-qu}) and the SL$(2,R)$ Casimir operator
(\ref{q-Cas-3}) can
be represented as
\begin{equation}\label{H-qu-g}
\mathbf{H} =\frac{1}{4}\,\left(P^2  +\frac{\hat g}{X^2} \right)\,,
\end{equation}
\begin{equation}\label{q-Cas-g}
\mathbf{T}^2 = {\textstyle\frac{1}{4}}\,\hat g - {\textstyle\frac{3}{16}}\,,
\end{equation}
where
\begin{equation}\label{hat-g}
\hat g \equiv 4\alpha^2{\textstyle\frac{1}{2}}\,\bar Z_k Z^{k}
\left({\textstyle\frac{1}{2}}\,\bar
Z_k Z^{k}+1\right) - 8\alpha Z^{(i} \bar Z^{k)} \Psi_{(i}\bar\Psi_{k)}- 2(1+2\alpha)
\langle\Psi_{i}\Psi^{i}\,\bar\Psi^{k} \bar\Psi_{k}\rangle + {\textstyle\frac{1}{4}}\,(1+2\alpha)^2\,.
\end{equation}
The operators (\ref{H-qu-g}) and (\ref{q-Cas-g}) formally look like those given in
the model of ~\cite{AFF}.
However, there is an essential difference. Whereas the quantity $\hat g$ is a
constant in
the model of ~\cite{AFF}, in our case $\hat g$ is an operator which takes fixed, but different,
constant values
on different components of the full wave function.

To find the quantum spectrum of (\ref{H-qu-g}) and (\ref{q-Cas-g}), we
make use of
the realization
\begin{equation}\label{bo-re-Z}
\bar Z_i=v^+_i, \qquad Z^i=  \partial/\partial v^+_i
\end{equation}
for the
bosonic operators $Z^k$ and $\bar{Z}_k$, as well as
the following realization of the odd operators $\Psi^i$, $\bar\Psi_i$
\begin{equation}\label{q-re-Psi}
\Psi^i=\psi^i, \qquad \bar\Psi_i= -{\textstyle\frac{1}{2}}\,
\partial/\partial\psi^i\,,
\end{equation}
where $\psi^i$ are complex Grassmann variables. Then, the wave function
is defined as
\begin{equation}\label{w-f}
\Phi=A_{1}+ \psi^i B_i +\psi^i\psi_i A_{2}\,.
\end{equation}

The full wave function is subjected to the same constraints~(\ref{D0-con})
as in the
bosonic limit
(we use the normal ordering for the even SU(2)-spinor operators, with all
operators
$Z^i$ standing on the right)
\begin{equation}\label{q-con}
D^0 \Phi=\bar Z_i Z^i \Phi=v^+_i\frac{\partial}{\partial
v^+_i}\,\Phi=c\,\Phi.
\end{equation}
Like in the bosonic limit, requiring the wave function $\Phi(v^+)$ to be
single-valued gives rise to the condition that
the constant $c$ is integer, $c\in \mathbb{Z}$. We take $c$ to be
positive in
order to have a correspondence with
the bosonic limit where $c$ becomes ${\rm SU}(2)$ spin. Then \p{q-con}
implies that
the wave function $\Phi(v^+)$
is a homogeneous polynomial in $v^+_i$ of the degree $c$:
\begin{equation}\label{w-f-d}
\Phi=A^{(c)}_{1}+ \psi^i B^{(c)}_i +\psi^i\psi_i A^{(c)}_{2} \,,
\end{equation}
\begin{equation}\label{A-irred}
A^{(c)}_{i^\prime} = A_{i^\prime,}{}_{k_1\ldots k_{c}}v^{+k_1}\ldots
v^{+k_{c}} \,,
\end{equation}
\begin{equation}\label{B-irred}
B^{(c)}_i = B^{\prime(c)}_i +B^{\prime\prime(c)}_i=v^+_i B^\prime_{k_1\ldots
k_{c-1}}v^{+k_1}\ldots v^{+k_{c-1}} + B^{\prime\prime}_{(ik_1\ldots
k_{c})}v^{+k_1}\ldots
v^{+k_{c}}\,.
\end{equation}
In~(\ref{B-irred}) we extracted ${\rm SU}(2)$ irreducible parts
$B^\prime_{(k_1\ldots
k_{c-1})}$ and $ B^{\prime\prime}_{(ik_1\ldots k_{c})}$ of the component wave
functions, with the ${\rm SU}(2)$
spins $(c-1)/2$ and $(c+1)/2$, respectively.

On the physical states~(\ref{q-con}), (\ref{w-f-d})  Casimir
operator~(\ref{qu-Cas-12}) takes the value
\begin{equation}\label{qu-Cas-ev}
\mathbf{C}_2= \alpha(1+\alpha)(c+1)^2/4 \, .
\end{equation}

On the same states, the Casimir operators  (\ref{q-Cas-3})-(\ref{q-Cas-1})
of the bosonic subgroups ${\rm SU}(1,1)$, ${\rm SU(2)}_R$ and ${\rm SU(2)}_L$
take the values given in the Table 1.\footnote{
Here we use that
$$
\Psi_i\Psi^i\, \bar\Psi^k\bar\Psi_k -
\Psi_i \bar\Psi^i  =\frac{1}{4} \left(\psi^i\psi_i\,
\frac{\partial}{\partial\psi_k}\frac{\partial}{\partial\psi^k} -
2\psi^i\frac{\partial}{\partial\psi^i} \right) \,,\qquad Z^{i} \bar Z^{k}\Psi_{(i}\bar\Psi_{k)}
= -\frac{1}{2} \left(v^{+i}\frac{\partial}{\partial v^+_j}
\psi_{(i}\frac{\partial}{\partial\psi^{j)}} \right) \,.
$$
Therefore, we have
$$
\Big(\Psi_i\Psi^i\, \bar\Psi^k\bar\Psi_k -
\Psi_i \bar\Psi^i \Big) \,\Phi= -\frac{1}{2} \psi^i B_i \,,\qquad
\Big(Z^{i} \bar Z^{k}\Psi_{(i}\bar\Psi_{k)} \Big)\,\Phi
= -\frac{1}{2} v^{+i}\frac{\partial}{\partial v^+_j}
\psi_{(i} B_{j)} = \frac{1}{4}
\psi^{i} \left[ (c+2) B^{\prime}_{i} -c \,B^{\prime\prime}_{i}\right]\,.
$$
}
\begin{table}[h]
\caption{The values of the Casimirs of the bosonic subgroups and $\frac{i}{4} \mathbf{Q}^{ai^\prime i}\mathbf{Q}_{ai^\prime i}$}
\label{tab1}
\begin{center}
\renewcommand{\arraystretch}{2}
\begin{tabular}{|c|c|c|c|c|}
\hline & $\mathbf{T}^2$ & $\mathbf{J}^2$ & $\mathbf{I}^2$ & $\frac{i}{4} \mathbf{Q}^{ai^\prime i}\mathbf{Q}_{ai^\prime i}$ \\ \hline
     $A^{(c)}_{k^\prime}$ & $\frac{\alpha^2(c+1)^2-1}{4}$ & $\frac{(c+1)^2-1}{4}$ &
$\frac{3}{4}$ & $1+\alpha$\\
\hline
     $B^{\prime(c)}_{k}$ & $\frac{\alpha^2(c+1)^2-2\alpha(c+1)}{4} $ & $\frac{(c+1)^2-2(c+1)}{4}$ & 0 & $\alpha(c+1)$ \\ \hline
     $B^{\prime\prime(c)}_{k}$ & $\frac{\alpha^2(c+1)^2+2\alpha(c+1)}{4}$ &
     $\frac{(c+1)^2+2(c+1)}{4}$& 0 & $-\alpha(c+1)$ \\ \hline
\end{tabular}\\
\end{center}
\end{table}
For different component wave functions, the quantum numbers $r_0, j$ and
$i$, defined by
$$
\mathbf{T}^2=r_0(r_0-1)\,, \qquad \mathbf{J}^2=j(j+1)\,,
\qquad \mathbf{I}^2=i(i+1)\,,
$$
take the values listed in the Table 2.
\begin{table}[h]
\caption{The ${\rm SU}(1,1)$, ${\rm SU(2)}_R$ and ${\rm SU(2)}_L$ quantum numbers}
\label{tab2}
\begin{center}
\renewcommand{\arraystretch}{2}
\begin{tabular}{|c|c|c|c|}
\hline & $r_0$ & $j$ & $i$ \\ \hline
     $A^{(c)}_{k^\prime}(x,v^+)$ & $\frac{|\alpha|(c+1)+1}{2}$ & $\frac{c}{2}$ &
$\frac{1}{2}$ \\
\hline
     $B^{\prime(c)}_{k}(x,v^+)$ & $\frac{|\alpha|(c+1)+1}{2} - \frac{1}{2}\,\rm{sign}(\alpha)$ & $\frac{c}{2}
-\frac{1}{2}$ & 0 \\ \hline
     $B^{\prime\prime(c)}_{k}(x,v^+)$ & $\frac{|\alpha|(c+1)+1}{2} + \frac{1}{2}\,\rm{sign}(\alpha)$ &
     $\frac{c}{2}+ \frac{1}{2}$& 0 \\ \hline
\end{tabular}\\
\end{center}
\end{table}
The fields $B^{\prime}_i$ and $B^{\prime\prime}_{i}$ form doublets
of SU(2)$_R$ generated by $\mathbf{J}^{ik}\,$, whereas the
component fields $A_{i^\prime}=(A_{1},A_{2})$ form a doublet of ${\rm
SU(2)}_L$
generated by
$\mathbf{I}^{i^\prime k^\prime}$. If the super-wave function (\ref{w-f})
is bosonic
(fermionic),
the fields $A_{i^\prime}$ describe bosons (fermions), whereas the fields
$B^{\prime}_i$, $B^{\prime\prime}_{i}$
present fermions (bosons). It is easy to check that the relation
\p{id-Cas} is valid in all cases.

Each of the component wave functions  $A_{i^\prime}$, $B^{\prime}_i$,
$B^{\prime\prime}_{i}$  carries an
infinite-dimensional unitary representation of the discrete series of the
universal
covering group of the one-dimensional conformal group SU(1,1). Such representations are characterized
by positive
numbers $r_0$~\cite{Barg,Per}
(for the unitary representations of SU(1,1) the constant $r_0 >0$ must be
(half)integer).
Basis functions of these representations are eigenvectors of the compact SU(1,1) generator
$$
\mathbf{R}={\textstyle\frac{1}{2}}\,\left(a^{-1}\mathbf{K}+
a\mathbf{H}\right),
$$
where $a$ is a constant of the length dimension. These eigenvalues are
$r=r_0 +n$,
$n\in \mathbb{N}$~\cite{Barg,Per,AFF}.

Using the expressions (\ref{H-qu}), (\ref{q-Cas-3})-(\ref{q-Cas-1}) and the values
of Casimirs from the Table 1, we
can write the Hamiltonian in the unified form:
\begin{equation}\label{H-com}
\mathbf{H} =\frac{1}{4}\,\left(P^2  +\frac{l(l+1)}{X^2}\right)
\end{equation}
where  the constant $l$ takes, on the separate wave functions, the values listed in the Table 3.
\begin{table}[h]
\caption{Values of the constant $l$}
\label{tab3}
\begin{center}
\renewcommand{\arraystretch}{1.8}
\begin{tabular}{|c|c|c|c|}
\hline & $l$ \\
\hline
     $A^{(c)}_{k^\prime}(x,v^+)$ & $|\alpha|(c+1) - \frac{1}{2}$  \\ \hline
     $B^{\prime(c)}_{k}(x,v^+)$ & $|\alpha|(c+1) - \frac{1}{2}-\rm{sign}(\alpha)$ \\ \hline
     $B^{\prime\prime(c)}_{k}(x,v^+)$ & $|\alpha|(c+1) - \frac{1}{2}+\rm{sign}(\alpha)$ \\ \hline
\end{tabular}\\
\end{center}
\end{table}

In the above quantization, we took into account all the conditions implied by the initial classical system.
Due to the presence of additional invariant spaces in the enveloping algebra, we may try to impose
additional conditions on the wave function, e.g.
\begin{equation}\label{L-cond}
\mathbf{L} \, \Phi= 0 \,
\end{equation}
where $\mathbf{L}$ was defined in \p{L}.
As a result, we could expect to obtain more restricted
spectrum at certain values of the parameters $\alpha$ and $c$.
Regrettably, this conjecture fails: in order to preserve the superconformal  $D(2,1;\alpha)$
covariance, we are led to assume
that all operators from the set (\ref{L})-(\ref{L-6}), on equal footing with $\mathbf{L}\,$, annihilate
the physical states, and these restrictions prove to be too strong.
It is an open question whether the constraints of this kind
could have a non-trivial solution in some other $D(2,1;\alpha)$ invariant superconformal mechanics models.

Let us focus on some peculiar properties of the $D(2,1;\alpha)$ quantum mechanics
constructed.

As opposed to the standard ${\rm SU}(1,1|2)$ superconformal mechanics~\cite{IKL,AIPT,Wyl}, the
construction presented here essentially uses the variables $z_i$ (or $v^+_i$)
parametrizing the two-sphere $S^2$, in addition to the standard (dilatonic) coordinate
$x$.

The presence of additional ``(iso)spin'' $S^2$ variables in our construction leads to a
richer quantum spectrum. Besides, the relevant wave functions involve representations of the
two independent
SU(2) groups, in contrast to the ${\rm SU}(1,1|2)$ models of \cite{IKL,AIPT,Wyl,GLP} where only the SU(2)
realized on fermionic variables really matters.

Also, in a contradistinction to the previously considered models
(and in the same way as in our previous paper \cite{FIL-09} devoted
to the particular $\alpha{=}-1/2$ case), there naturally
appears a
quantization of the conformal coupling constant which is expressed as a SU(2)
Casimir operator,
with both integer and half-integer eigenvalues. This happens already in the bosonic
sector of the model,
and is ensured by the $S^2$ variables.\footnote{Note that the strength of the conformal
potential is related to the strength of the WZ term
and so is quantized also in the ${\cal N}{=}4$
superconformal mechanics associated with the ${\bf (3, 4,1)}$ multiplet (without non-dynamical
$S^2$ variables)\cite{IKLech}. However, no direct relation
between these parameters and SU(2) Casimirs appears in this case.}

Note that the variables $v^+_i$ in the expansions (\ref{A-irred}) and (\ref{B-irred})
can be identified with a half of the target space
harmonic-like variables $v^\pm_i$ (though without the standard constraint $v^{+ i}v_i^- \sim
const$). Within a different quantization scheme used e.g. in \cite{ABS},
we would have even more literal harmonic interpretation of the bosonic isospinor variables.
In both schemes, the $S^2$ constraint \p{con} is not explicitly solved before quantization, it is imposed
on the wave functions as in \p{q-con}. An alternative quantization scheme would be to deal with an explicit
parametrization of the two-spere $S^2$, e.g. the stereographic projection parametrization~\cite{IM}
or the parametrization by the Euler angles $\beta$ and $\gamma$ as in \p{bose1},
and then to apply the canonical
methods (Gupta-Bleuler quantization or Dirac procedure).\footnote{One more approach
is to quantize in the oscillator variables \cite{FJ,Poly}.}
An important role in this case is played by the requirement of the square-integrability
of the wave function on $S^2$, which substitutes the constraint \p{q-con} of the parametrization-independent
quantization schemes. As follows from the consideration in \cite{Poly,IM}, this demand ensures
the wave function to contain unitary representations of SU(2).
General issues of the canonical quantization of Chern--Simons mechanics were addressed in~\cite{RPS}.

\subsection{Comment on the ${\rm SU}(1,1|2)$ case}

Let us here focus on some peculiarities of the case of ${\rm SU}(1,1|2)$ superconformal symmetry.

In the case of $\alpha{=}-1$ one has $D(2,1;\alpha{=}-1)\simeq {\rm
SU(1,1|2)}{\subset\!\!\!\!\!\!\times}{\rm SU(2)}_L$, and thus our model is invariant under
${\rm SU}(1,1|2)$ superconformal group and an outer automorphism group ${\rm SU(2)}_L$
acting only on the fermions.
In general, the supergroup ${\rm SU}(1,1|2)$ is known to admit a non-vanishing
central charge which breaks this second R-symmetry
SU(2) group down to U(1)~\cite{IKL}\footnote{The quotient of the general
${\rm SU}(1,1|2)$ over the central charge generator is sometimes
denoted as ${\rm PSU}(1,1|2)\,$.}. Thus, if we require our model to be
invariant under ${\rm SU(2)}_L$ (as in the case of generic $\alpha$)
the corresponding ${\rm SU}(1,1|2)$ algebra cannot include a central charge.

There arises the question as to whether a different version of the ${\cal N}{=}4$
superconformal mechanics model with spin variables exists, such that it possesses ${\rm
SU}(1,1|2)$ symmetry with a non-vanishing central charge. The answer is affirmative,
and it can be derived from the results of refs.~\cite{IKL,Wyl,GLP}.

When only ${\rm SU}(1,1|2)$ symmetry is required,
while ${\rm SU(2)}_L$ symmetry is allowed to be broken, the
constraints (\ref{cons-X-g-V}) and (\ref{cons-X-g}) for the even real superfield
$\mathscr{X}$ can be weakened~\cite{IKL} by adding nonzero constants in their right-hand sides.
The simplest choice is
the following set of the constraints
\begin{equation}  \label{cons-X-rel1}
{\rm (a)} \quad D^iD_i \,\mathscr{X}=0\,, \;\; \bar D_i\bar D^i \,\mathscr{X}=0\,; \qquad  {\rm (b)} \quad
[D^i,\bar D_i]\, \mathscr{X}=m
\end{equation}
where $m$ is a constant. The solution of the constraints (\ref{cons-X-rel1}a) is a sum
of~(\ref{sing-X0-WZ}) and additional term $-\frac{1}{4}\theta\bar\theta A$, where $A$ is
some undefined constant. The constraint (\ref{cons-X-rel1}b) serves to fix this constant to be $m$. Then the
action~(\ref{4N-X}) (with $\alpha{=}{-}1$) will give rise to additional contributions to the physical
component Lagrangian~(\ref{4N-ph}), such that they are  proportional
to $m^2/x^2$ and $m\psi\bar\psi /x^2$~\cite{IKL}.
These additional terms appear in the Hamiltonian, and they are induced by the appropriate new terms
in the Noether supercharges.
Comparing these modified ${\rm SU}(1,1|2)$ generators with those
given in~\cite{Wyl,GLP}, one can see that they correspond just to
the ${\rm SU}(1,1|2)$ algebra with a central charge proportional to $m$.

More detailed analysis of the U(2) spin ${\cal N}{=}4$ superconformal mechanics in which
the even real superfield $\mathscr{X}$ is subjected to the constraints (\ref{cons-X-rel1})
with $m\neq 0$ will be given elsewhere. An interesting new feature of such
a model is the presence of two complementary mechanisms of generating the conformal potential $\sim x^{-2}$:
the on-shell one via coupling to the auxiliary superfields
${\mathcal{Z}}^+$ as in the case of generic $\alpha$, and the off-shell one based on the deformed
constraints (\ref{cons-X-rel1}) and a non-zero central charge in the ${\rm SU}(1,1|2)$ algebra.
It should be stressed that
such a modification of the constraints is admissible only in the case of $\alpha{=}{-}1\,$;
at any other value of $\alpha$ (not belonging to the equivalence class of the choice $\alpha{=}{-}1$)
the superconformal invariance requires the constants in the right-hand sides of the constraints to vanish.

\setcounter{equation}0
\section{Summary and outlook}

In this paper we presented a new version of ${\cal N}{=}4$ mechanics
with $D(2,1;\alpha)$ superconformal symmetry. It is obtained as the
one-particle reduction of the many-particle Calogero-type systems
proposed in~\cite{FIL}. This system generalizes the ${\rm OSp}(4|2)$
superconformal mechanics constructed in our previous work~\cite{FIL-09},
and it shares many characteristic features of the latter. In
the bosonic sector it involves two complex fields (world-line harmonics)
parametrizing the first Hopf map $S^3\rightarrow S^2$.

Due to the presence of spin variables in the superconformal mechanics,
the quantum spectrum involves diverse $D(2,1;\alpha)$ representations
characterized by the specific values of the Casimir operator~(\ref{qu-Cas}),
(\ref{qu-Cas-12}). In these representations, the particle states carry
representations of the bosonic subgroups ${\rm SU}(1,1)$, ${\rm SU(2)}_L$
and ${\rm SU(2)}_R$, the Casimirs of which are related to each other by
the constraint~(\ref{id-Cas-1}). This constraint is identically satisfied
for the particular realization of the $D(2,1;\alpha)$ generators pertinent
to our model.

The appearance of this constraint is related to the existence of some
invariant subspaces in the enveloping algebra of $D(2,1;\alpha)$.
We found that at generic $\alpha$ there exist more invariant subspaces
than for the degenerate case of $\alpha{=}-1/2$ corresponding to
${\rm OSp}(4|2)$~\cite{FIL-09}, where some invariant subspaces are identified.

The $D(2,1;\alpha)$ superconformal mechanics was considered here
for $\alpha{\neq}0 $. Formally, we can take the limit $\alpha\rightarrow0$
in the final relations, and we observe that the target harmonic degrees of
freedom decouple (see, e.g., (\ref{bose}), (\ref{fermi}) and
(\ref{2str-X})--(\ref{2str-A})). Nevertheless, the superconformal superfield
action of the $({\bf 1,4,3})$ multiplet is of a special form for $\alpha{=}0$,
so this case requires a separate study. Here we give a brief comment on the
construction of the superfield superconformal action at $\alpha{=}0\,$.

We note that $D(2,1;\alpha{\to}0)$ reduces to
${\rm SU(1,1|2)}{\subset\!\!\!\!\!\!\times}{\rm SU(2)}_R$.
The ``passive'' superconformal variation~(\ref{sc-1n}) of $\mathscr{X}$
disappears in this case, while the integration measure
$\mu_H$ is transformed as (see~(\ref{sc-me1}))
\begin{equation}  \label{sc-me2-1}
\delta^\prime \mu_H= -2i (\theta_k\bar\eta^k + \bar\theta^k\eta_k)\,\mu_H\,.
\end{equation}
As suggested in~\cite{IKLech,DI1}, in order to ensure the superconformal
invariance, it is necessary to modify the transformation
law of $\mathscr{X}$ and, therefore, of $\mathscr{V}$ in the following way,
\begin{equation}  \label{sc-X-m}
\delta_{mod}^\prime \mathscr{X}=2i(\theta_k\bar\eta^k + \bar\theta^k\eta_k)\, ,
\qquad\delta_{mod}^\prime\mathscr{V}=4i(\bar\eta^-\theta^+-\eta^-\bar\theta^+)
\,.
\end{equation}
Then the most general $D(2,1;\alpha{=}0)$ superconformal action for the
$({\bf 1, 4, 3})$ multiplet reads~\cite{DI1}
\begin{equation}\label{4N-X-0}
S^{\mathscr{X}}_{\alpha= 0}
 =  -{\textstyle\frac{1}{4}} \int \mu_H \,   e^{\,\mathscr{X}} +
 \int \mu^{(-2)}_A  c^{+2} \,\mathscr{V}
\, ,
\end{equation}
where $c^{+2}=c^{ij}u^+_iu^+_j$, and $c^{ij}$ are constant parameters.
The second FI term in~(\ref{4N-X-0})
is superconformal only at $\alpha{=}0\,$. It yields a conformal potential for
the dilaton field with a strength $\sim c^{ik}c_{ik}\,$, breaks the decoupled
SU(2)$_R$ down to U(1) and induces a central charge
$\sim c^{ik}$ in ${\rm SU}(1,1|2)\,$. Actually, this action is dual to
the $\alpha{=}{-}1$ action for $\mathscr{X}$
with the modified constraints (\ref{cons-X-rel1})~\cite{ikrpa}:
the duality interchanges SU(2)$_L$ with SU(2)$_R$ and also
$\alpha$ with $-(1{+}\alpha)\,$.
However, the $D(2,1;\alpha{=}0)$ superconformal invariance is not compatible
with the presence of $\mathcal{V}$ in the WZ term of the action~(\ref{4N-WZ}),
still implying the transformation laws~(\ref{sc-1n}) for $\mathcal{Z}^+$
and for~$V^{++}\,$. As a consequence, the WZ term and the FI term of $V^{++}$
decouple from the $\mathscr{X}$ action:
\begin{equation}\label{4N-WZ-0}
S^{\mathcal{Z},V^{++}}_{\alpha= 0} =
 {\textstyle\frac{1}{2}}\int \mu^{(-2)}_A\,
\widetilde{\mathcal{Z}}{}^+ \mathcal{Z}^+
+  {\textstyle\frac{i}{2}}\,c\int \mu^{(-2)}_A \,V^{++} \,  .
\end{equation}
i.e.\ we loose any interaction between the superfields $\mathscr{X}$ and
$\mathcal{Z}^+$. This situation is quite analogous to what happens in
the ${\cal N}{=}1$ and ${\cal N}{=}2$ super Calogero models considered
in~\cite{FIL}, where the center-of-mass supermultiplet~$\mathscr{X}$
decouples from the WZ and gauge supermultiplets.
Note that in the many-particle ${\cal N}{=}4$ super Calogero models
the (matrix) $\mathscr{X}$ supermultiplet will still interact with
the (column) $\mathcal{Z}$ supermultiplet via the gauge supermultiplet even
in the $\alpha{=}0$ case.

Based on the duality just mentioned between the cases of $\alpha{=}0$
and $\alpha{=}{-}1$, one may expect that in the $\alpha{=}0$ case
the interaction of the superfield~$\mathscr{X}$ with the U(2) spin variables
can still be gained by placing the latter into a ``mirror''
$({\bf 4, 4,0 })$ multiplet, for which the ${\rm SU(2)}_R$ and ${\rm SU(2)}_L$
R-symmetry groups switch their roles.
In this context, it is worth noting that the bi-harmonic ${\cal N}{=}4$
approach~\cite{IN} achieves a unified description of systems with
$D(2,1;\alpha)$ and $D(2,1;{-}1{-}\alpha)$ invariance. It allows one to
naturally incorporate mirror counterparts for all ${\cal N}{=}4$
supermultiplets with four fermions. Hence, it may provide an extension
of the $D(2,1;\alpha)$ superconformal models considered here,
by adding such extra supermultiplets. Upon quantization,
the mirror $({\bf 4, 4, 0})$ auxiliary multiplets would produce
a second family of target harmonic-like U(2)~variables.

For the remainder of this outlook and as a continuation of the discussion
in the Introduction,
let us illustrate how the models considered in this paper and
in~\cite{FIL,FIL-09} could be inscribed into the context of $D{=}5$
extreme black-hole quantum mechanics.

The motion of a test particle with mass~$m$ near
the horizon of an extremal Tangherlini black hole of charge~$Q$
(a straightforward $D{=}5$ generalization of the $D{=}4$ extremal
Reissner-Nordstr\"{o}m solution) is described by the simple action~\cite{MS}
\begin{equation}\label{T-ac}
S = {\textstyle\frac{mQ^2}{2}}\int\,dt\,|\dot{\vec{y}}|^2 \,,
\end{equation}
where $\vec{y}$ are the coordinates of Euclidean four-space which are
related to the isotropic near-horizon black-hole coordinates $\vec{x}$
via $\vec{y}=\vec{x}/|\vec{x}|^2$.

Making a polar decompostiion of the 4-vector $\vec{y}$ into a radial part
$\rho=|\vec{y}|$ and an $S^3$ angular part, we rewrite the action~(\ref{T-ac})
in first-order form as
\begin{equation}\label{kin}
S = \int\,\left[p_\rho d\rho+\vec{J}{\cdot}\vec{\omega} -
dt\, {\textstyle\frac{1}{2mQ^2}} \left(
p_\rho^2+{\textstyle\frac{4 \vec{J}{\cdot}\vec{J}}{\rho^2}}\right)\right].
\end{equation}
Here, $\omega_i$ are the invariant one-forms on $S^3\sim{\rm SU(2)}\,$,
parametrized by the Euler angles
($0{\leq}\gamma{\leq}\pi$, $0{\leq}\beta{\leq}2\pi$, $0{\leq}\phi{<}4\pi$):
\begin{equation}\label{om-3}
\omega_1 =-\sin\!\phi\,d\gamma +\cos\!\phi\,\sin\!\gamma\,d\beta \,,\quad
\omega_2 = \cos\!\phi\,d\gamma +\sin\!\phi\,\sin\!\gamma\,d\beta \,,\quad
\omega_3 = d\phi +\cos\!\gamma\,d\beta\,.
\end{equation}
In the Hamiltonian approach, the quantities $\vec{J}$ generate
some ${\rm SU(2)}$ invariance~\cite{GGHPR}.
It is easy to see that the action (\ref{T-ac}) is indeed reproduced
by eliminating $p_\rho$ and $\vec{J}$ in \p{kin}
by their algebraic equations of motion. Firstly, we obtain the action
\begin{equation}\label{ac-sim}
S = {\textstyle\frac{mQ^2}{2}} \int dt \left[\dot\rho \dot\rho+
\rho^2{\textstyle\frac{1}{4}}\vec{\omega}_t{\cdot}\vec{\omega}_t \right]
\qquad\textrm{where}\qquad \vec{\omega}=\vec{\omega}_tdt\,.
\end{equation}
However, ${\textstyle\frac{1}{4}}\vec{\omega}{\cdot}\vec{\omega}$
is precisely the $S^3$ metric~\cite{GMT-2,GGHPR}.
Therefore secondly, the action takes the form
\begin{equation}
S = {\textstyle\frac{mQ^2}{2}} \int dt \left[\dot\rho \dot\rho+
\rho^2\dot{\vec{n}}{\cdot}\dot{\vec{n}} \right]
\qquad\textrm{where}\qquad |\vec{n}|=1\,.
\end{equation}
This is just (\ref{T-ac}) with $\vec{y}=\rho\,\vec{n}$.

Performing in \p{kin} a reduction with respect to
the variables $\vec{J}$~\cite{Per-b},
\begin{equation}\label{red}
J_1=J_2=0\,,\qquad J_3=a=\textrm{const}\,,
\end{equation}
and identifying $\rho=bx$, $p_\rho=b^{-1}p_x$, $a=-c/2\,$,
where $b^2=\frac{2}{m Q^2}$,
we obtain the one-particle bosonic limit (\ref{bose1})
of the action~(\ref{4N-gau-matrix}) at $|\alpha|=1$.

The fact that just this particular value of $\alpha$ comes out
is not surprising because the action~(\ref{T-ac}) was obtained in~\cite{MS}
as the bosonic limit of the ${\rm SU}(1,1|2)$ superconformal model.
It is interesting that the action~(\ref{bose1}) at {\it arbitrary\/}
non-zero value of~$\alpha$ can still be reproduced
by the same reduction~\p{red} from a deformation of the action~(\ref{T-ac})
(or, equivalently, of~(\ref{kin})).

This can be done in two different ways.
One option is to substitute \ $4 (J_1J_1+J_2J_2+\alpha^2J_3J_3)/{\rho^2}$ \
for \ ${4 \vec{J}{\cdot}\vec{J}}/{\rho^2}$ \
in the last term of~(\ref{kin}).
The action~(\ref{ac-sim}) deformed in this way involves the metric \
${\textstyle\frac{1}{4}}({\omega}_{t1} {\omega}_{t1} +{\omega}_{t2}
{\omega}_{t2}+ \alpha^{-2}{\omega}_{t3} {\omega}_{t3})$ \
instead of \ ${\textstyle\frac{1}{4}}\vec{\omega}_t{\cdot}\vec{\omega}_t$.
Such a system describes the particle motion on a squashed 3-sphere,
with $\alpha^{-2}$ as the squashing parameter.
This model may bear a tight relation to $D{=}5$ rotating black holes,
whose horizon is known to be a squashed 3-sphere~\cite{BMPV,GMT-2,Town}.
The O(4)~symmetry of~\p{T-ac} is broken to~O(3) in this situation.

Another possibility is to replace ${4 \vec{J}{\cdot}\vec{J}}/{\rho^2}$ in
the last term of~(\ref{kin}) by ${4 \alpha^2\vec{J}{\cdot}\vec{J}}/{\rho^2}$.
The Lagrangian in~(\ref{ac-sim}) is then deformed into \ $[\dot\rho \dot\rho
+\alpha^{-2}\rho^2{\textstyle\frac{1}{4}}\vec{\omega}_t{\cdot}\vec{\omega}_t]$.
This system describes particle motion on a 4-dimensional cone~$C(S^3)$ over
the round sphere~$S^3$ of radius~$\alpha^{-2}$ as the base~\cite{GibR,IKLech}.
This cone is conformally flat and exhibits O(4)~isometry at
any~$\alpha{\neq}0\,$,
including the values $\alpha{=}{\pm}1$ which correspond to the action~\p{T-ac}.

In both cases, the reduction~(\ref{red}), performed in the relevant
counterparts of the action~(\ref{kin}), exactly yields
our action~(\ref{bose1}). It is amusing that the parameter $\alpha$ acquires
a nice geometric meaning within such a framework.

\section*{Acknowledgements}

\noindent We acknowledge support from a grant of the Heisenberg-Landau Programme, RFBR
grants 08-02-90490, 09-02-01209, 09-01-93107 (S.F. \&\
E.I.) and a DFG grant, project No.~436 RUS/113/669 (E.I. \&\ O.L.).
SF would like to thank KEK, Tsukuba,
Yukawa Institute for Theoretical Physics, Kyoto and
the Institute of Theoretical Physics at Leibniz University
of Hannover for the warm hospitality at the final stage of this study.

\renewcommand\theequation{A.\arabic{equation}} \setcounter{equation}0
\section*{Appendix A: \quad  Toy model with ${\cal N}{=}2$ supersymmetry}

Here we consider ${\cal N}{=}2$ supersymmetric model describing a ``matter'' supermultiplet coupled to U(1)
gauge background. Matter is represented  by two chiral superfields $ {Z}^k
(t_{\!\scriptscriptstyle{L}}, \theta) $, $ \bar {Z}_k (t_{\!\scriptscriptstyle{R}},
\bar\theta) = ({Z}^k)^+$, $ t_{\!\scriptscriptstyle{L,R}}=t\pm i\theta\bar\theta$,
satisfying irreducible conditions $ \bar D {Z}^k=0 $, $D \bar {Z}_k=0$, $k=1,2$. Here, the
covariant spinor derivatives are
$$
D = \partial_{\theta} +i\bar\theta\partial_{t}\,, \quad \bar D = -\partial_{\bar\theta}
-i\theta\partial_{t}\,, \quad \{D, \bar D \} = -2i \partial_{t}\,.
$$
The gauge prepotential is a real superfield $ V(t, \theta,\bar\theta)$, $(V)^+ =V\,$. The
action has the following form
\begin{equation}\label{N2-Cal}
S = \int dt d^2\theta \,\Big[\, \bar {Z}_k\, e^{2V}\! {Z}^k + c\, V \,\Big]\,.
\end{equation}
It is invariant under the local $U(1)$ transformations:
\begin{equation}\label{Un-tran-s}
{Z}^k \rightarrow \, e^{-i\Lambda} {Z}^k \,, \qquad \bar {Z}_k \rightarrow \,
e^{i\bar\Lambda} \bar {Z}_k\, \,, \qquad V \rightarrow \, V+\frac{i}{2}\,\left(\Lambda
-\bar\Lambda\right)
\end{equation}
where $ \Lambda (t_{\!\scriptscriptstyle{L}}, \theta) $, $\bar\Lambda
(t_{\!\scriptscriptstyle{R}}, \theta) = (\Lambda)^+ $ are chiral and antichiral
superfield gauge parameters.

Supersymmetry transformations of a general ${\cal N}{=}2$ superfield $F$ are defined by
\begin{equation}\label{SUSY-gen}
\delta F = -(\delta t\partial_t +\delta \theta\partial_\theta+ \delta
\bar\theta\partial_{\bar\theta})\,F = -(\varepsilon\, Q - \bar\varepsilon\, \bar Q)\,F
\end{equation}
where the generators of SUSY transformations are
$$
Q = \partial_{\theta} -i\bar\theta\partial_{t}\,, \quad \bar Q = -\partial_{\bar\theta}
+i\theta\partial_{t}\,.
$$
Component contents of the superfields defined above are
\begin{equation}\label{com-V}
{Z}^k = z^k + 2i\theta \phi^k + i\theta\bar\theta \dot z^k \,, \qquad \bar{Z}_k = \bar z_k
+ 2i\bar\theta \bar\phi_k - i\theta\bar\theta \dot{\bar z}_k\,, \qquad V = v + \theta \chi
- \bar\theta \bar\chi + \theta\bar\theta A \,,
\end{equation}
where $\phi^k$, $\bar\phi_k=(\overline{\phi^k})$ and $\chi$, $\bar\chi=(\overline{\chi})$
are fermionic fields. For the component fields the transformations~(\ref{SUSY-gen}) yield
\begin{equation}\label{SUSY-Z}
\delta z^k = -2i\varepsilon\phi^k\,,\qquad  \delta \bar z_k =
-2i\bar\varepsilon\bar\phi_k\, ,\qquad \delta\phi^k = -\bar\varepsilon  \dot z^k\, ,\qquad
\delta\bar\phi_k = - \varepsilon\dot{\bar z}_k\, ,
\end{equation}
\begin{equation}\label{SUSY-V}
\delta v = -\varepsilon\chi + \bar\varepsilon\bar\chi\, ,\qquad \delta\chi =
-\bar\varepsilon( A +i \dot v)\, ,\qquad \delta\bar\chi = -\varepsilon( A -i \dot v)\,
,\qquad \delta A = -i(\varepsilon\dot\chi + \bar\varepsilon\dot{\bar\chi})\, .
\end{equation}

Let us consider the action~(\ref{N2-Cal}) in the WZ gauge,
\begin{equation}\label{WZ-2}
V (t, \theta,\bar\theta) = \theta\bar\theta A (t)\,, \quad e^{2V} = 1+ 2\theta\bar\theta A\,.
\end{equation}
It takes the form ($\int d^2\theta\, (\theta\bar\theta) =1 $)
$$
S^{\scriptscriptstyle{WZ}} =  {\displaystyle\int} dt \,\Bigg[ \,i (\bar z_k \nabla z^k -
\nabla \bar z_k z^k) + c\,A \, - 4\bar\phi_k\phi^k\,\Bigg], \nonumber
$$
where $\nabla z$ and $\nabla \bar z$ are the gauge-covariant derivatives,
\begin{equation}\label{cov-der-Psi}
\nabla z^k = \dot z^k -i A z^k\,, \qquad \nabla \bar z_k = \dot {\bar z}_k +i A \bar z_k\,.
\end{equation}
The action~(\ref{cov-der-Psi}) is invariant under the residual local $U(1)$ transformations
\begin{equation}\label{Un-tran-res}
\delta{z}^k =-i\lambda {z}^k \,, \qquad \delta\bar {z}_k = i\lambda \bar {z}_k\, \,, \qquad
\delta A = -\dot\lambda\,,
\end{equation}
where $ \lambda (t) $ is the $d{=}1$ gauge parameter.

Supersymmetry transformations~(\ref{SUSY-Z})-(\ref{SUSY-V}) do not preserve the WZ gauge conditions
$v=0$, $\chi =0$, $\bar\chi =0\,$, and we are led to modify these transformations by a field-dependent
compensating gauge transformation with the parameter
$$
\Lambda = -2i\theta\bar\varepsilon A\,, \qquad \bar\Lambda = -2i\bar\theta\varepsilon A\,.
$$
Then the supersymmetry transformations leaving invariant the action~(\ref{cov-der-Psi}) are given by
\begin{equation}\label{SUSY-WZ-Z1}
\delta^{\scriptscriptstyle{WZ}} z^k = -2i\varepsilon\phi^k\,,\qquad
\delta^{\scriptscriptstyle{WZ}} \bar z_k = -2i\bar\varepsilon\bar\phi_k\, ,\qquad
\delta^{\scriptscriptstyle{WZ}}\phi^k = -\bar\varepsilon  \nabla z^k\, ,\qquad
\delta^{\scriptscriptstyle{WZ}}\bar\phi_k = -\varepsilon\nabla{\bar z_k}\, ,
\end{equation}
\begin{equation}\label{SUSY-A1}
\delta^{\scriptscriptstyle{WZ}} A = 0\, .
\end{equation}
Let us study the closure of these transformations. On the fields
$z^k$ we have
\begin{equation}\label{SUSY-WZ-z}
\left(\delta^{\scriptscriptstyle{WZ}}_1\delta^{\scriptscriptstyle{WZ}}_2-
\delta^{\scriptscriptstyle{WZ}}_2\delta^{\scriptscriptstyle{WZ}}_1\right) z^k =
2i\left(\varepsilon_1\bar\varepsilon_2-\varepsilon_2\bar\varepsilon_1\right) \nabla z^k=
2ia_{12}\dot z^k-i\lambda_{12}z^k\, ,
\end{equation}
where
\begin{equation}\label{a-la}
a_{12}= \varepsilon_1\bar\varepsilon_2-\varepsilon_2\bar\varepsilon_1  \,, \qquad
\lambda_{12} =
2i\left(\varepsilon_1\bar\varepsilon_2-\varepsilon_2\bar\varepsilon_1\right)A\,.
\end{equation}
Thus, the r.h.s. of~(\ref{SUSY-WZ-z}) is the time translation with the parameter $a_{12}$
accompanied by a residual gauge transformation with the parameter $\lambda_{12}$. Clearly, the closure on the
gauge field $A(t)$ should be the same. We find
\begin{equation}\label{SUSY-WZ-A}
\delta^{\scriptscriptstyle{WZ}}_{12} A = 2ia_{12}\dot A-\dot\lambda_{12}\, = 0,
\end{equation}
in agreement with~(\ref{SUSY-A1}).

On shell, after eliminating the auxiliary fields $\phi$, $\bar\phi$ in the action~(\ref{cov-der-Psi}),
\begin{equation}\label{eq-chi}
\phi^k=0\,, \qquad \bar\phi_k=0\,,
\end{equation}
the action \p{cov-der-Psi} and the supersymmetry transformations ~(\ref{SUSY-WZ-Z1}), (\ref{SUSY-A1}) become
\begin{equation}\label{N2Cal-com}
S^{\scriptscriptstyle{WZ}} =  {\displaystyle\int} dt \,\Bigg[\, i (\bar z_k \nabla z^k -
\nabla \bar z_k z^k) + c\,A\,\Bigg],
\end{equation}
\begin{equation}\label{SUSY-WZ-Z2}
\tilde\delta^{\scriptscriptstyle{WZ}} z^k = 0\,,\qquad
\tilde\delta^{\scriptscriptstyle{WZ}} \bar z_k = 0\, , \qquad
\tilde\delta^{\scriptscriptstyle{WZ}} A = 0\, .
\end{equation}
Taking into account the equations of motion
$$
\nabla z^k = \nabla \bar{z}_k = 0\,,
$$
these on-shell transformations close on the time translations and gauge transformation like their
off-shell counterparts~(\ref{SUSY-WZ-Z1}), (\ref{SUSY-A1}).

The structure of the component ${\cal N}{=}4$ supersymmetry transformations in the WZ gauge
in our $D(2,1;\alpha)$ superconformal mechanics model is basically the same as in the toy model just
considered.

\renewcommand\theequation{B.\arabic{equation}} \setcounter{equation}0
\section*{Appendix B: \quad  Time reversal in mechanics}

Let us consider the simple mechanical model with the Lagrangian
\begin{equation}\label{L1-t}
L_1 = \dot x^2 -i ( \bar\psi \dot\psi -\dot{\bar\psi} \psi ) -i ( \bar z \dot z -\dot{\bar
z} z )- U(x,\psi,\bar\psi,z,\bar z)\, .
\end{equation}
The canonical momenta are \footnote{We use the notations which are related
to those in~\cite{Cas} through a redefinition. In particular, we define the fermionic momenta as right
derivatives of the Lagrangian.}
\begin{equation}\label{mom-t}
p = 2\dot x,\qquad p_{\psi} =-i \bar\psi,\quad p_{\bar\psi} =-i \psi ,\qquad p_{z} =-i \bar
z ,\quad p_{\bar z} =i z
\end{equation}
with Poisson brackets
\begin{equation}\label{PB-mom-t}
[x, p]_{{}_P}= 1, \qquad [z, p_{z}]_{{}_P}= [\bar z, p_{\bar z}]_{{}_P}=1, \qquad \{\psi,
p_{\psi}\}_{{}_P}= \{\bar \psi, p_{\bar \psi}\}_{{}_P}=1\,.
\end{equation}
 Therefore, the Hamiltonian is
\begin{equation}\label{Ham-t}
H = p\dot x+ p_{\psi} \dot\psi+ p_{\bar\psi} \dot{\bar\psi} + p_{z} \dot z+ p_{\bar z}
\dot{\bar z} - L = {\textstyle\frac{1}{4}}\,p^2 +U.
\end{equation}
The definition~(\ref{mom-t}) implies second-class constraints
\begin{equation}\label{con-t}
G_{\psi}= p_{\psi} +i \bar\psi\approx0,\quad G_{\bar\psi}=p_{\bar\psi} +i \psi\approx0
,\qquad G_{z}= p_{z} +i \bar z\approx0 ,\quad G_{\bar z}=p_{\bar z} -i z\approx0;
\end{equation}
$$
\{G_{\psi}, G_{\bar\psi}\}_{{}_P}=2i\,,\qquad [G_{z}, G_{\bar z}]_{{}_P}=2i\,.
$$
Introducing Dirac brackets
\begin{eqnarray}
[A, B\}_{{}_D}=[A, B\}_{{}_P} &+&{\textstyle\frac{i}{2}}[A, G_{\psi}\}_{{}_P}[G_{\bar\psi},
B\}_{{}_P} +{\textstyle\frac{i}{2}}[A, G_{\bar\psi}\}_{{}_P}[G_{\psi}, B\}_{{}_P} \nonumber\\
&-&{\textstyle\frac{i}{2}}[A, G_{z}\}_{{}_P}[G_{\bar z}, B\}_{{}_P}
+{\textstyle\frac{i}{2}}[A, G_{\bar z}\}_{{}_P}[G_{ z}, B\}_{{}_P} \nonumber
\end{eqnarray}
we obtain
\begin{equation}\label{DB-t}
[x, p]_{{}_D}= 1, \qquad [z, \bar z]_{{}_D}={\textstyle\frac{i}{2}}, \qquad \{\psi, \bar
\psi\}_{{}_D}={\textstyle\frac{i}{2}}\,.
\end{equation}
Then, passing to quantum theory, we obtain the following operator algebra
\begin{equation}\label{QB-t}
[X, P]= i, \qquad [Z, \bar Z]=-{\textstyle\frac{1}{2}}, \qquad \{\Psi, \bar \Psi\}
=-{\textstyle\frac{1}{2}}\,.
\end{equation}

The time-reversed system is described by the Lagrangian\footnote{To be more precise, under the time reversal we also need to change the
sign of the overall normalization constant before the invariant action since the integral $\int dt$ changes its sign.}
\begin{equation}\label{L2-t}
L_2 = \dot x^2 +i ( \bar\psi \dot\psi -\dot{\bar\psi} \psi ) +i ( \bar z \dot z -\dot{\bar
z} z )- U(x,\psi,\bar\psi,z,\bar z)\, .
\end{equation}
Performing the same procedure as above we obtain that the system~(\ref{L2-t}) has the same
Hamiltonian~(\ref{Ham-t}), but different Dirac brackets
\begin{equation}\label{DB-t2}
[x, p]_{{}_D}= 1, \qquad [z, \bar z]_{{}_D}=-{\textstyle\frac{i}{2}}, \qquad \{\psi, \bar
\psi\}_{{}_D}=-{\textstyle\frac{i}{2}}
\end{equation}
which yield
\begin{equation}\label{QB-t2}
[X, P]= i, \qquad [Z, \bar Z]={\textstyle\frac{1}{2}}, \qquad \{\Psi, \bar \Psi\}
={\textstyle\frac{1}{2}}\,.
\end{equation}

Comparing \p{QB-t2} with \p{QB-t}, we observe that
the former turns into the latter after redefining
$$
\bar Z=-\left( Z\right)^+\,,\qquad \bar \Psi=-\left( \Psi\right)^+\,.
$$


\bigskip


\begin{thebibliography}{96}
\addtolength{\itemsep}{-5pt}

\bibitem{CDKKTP}
P.\,Claus, M.\,Derix, R.\,Kallosh, J.\,Kumar, P.K.\,Townsend, A.\,Van\,Proeyen,
{\it Black holes and superconformal mechanics},
Phys. Rev. Lett. {\bf 81} (1998) 4553, {\tt arXiv:hep-th/9804177}.

\bibitem{FIL}
S.\,Fedoruk, E.\,Ivanov, O.\,Lechtenfeld,
{\it Supersymmetric Calogero models by gauging},
Phys. Rev. {\bf D79} (2009) 105015, {\tt arXiv:0812.4276\,[hep-th]}.

\bibitem{DI}
F.\,Delduc, E.\,Ivanov,
{\it Gauging N=4 supersymmetric mechanics},
Nucl. Phys. {\bf B753} (2006) 211, {\tt arXiv:hep-th/0605211}.

\bibitem{GT}
G.W.\,Gibbons, P.K.\,Townsend,
{\it Black holes and Calogero models},
Phys. Lett. {\bf B454} (1999) 187, {\tt arXiv:hep-th/9812034}.

\bibitem{MS}
J.\,Michelson, A.\,Strominger,
{\it The geometry of (super)conformal quantum mechanics},
Commun. Math. Phys. {\bf 213} (2000) 1, {\tt arXiv:hep-th/9907191};
{\it Superconformal multi-black hole quantum mechanics},
JHEP {\bf 9909} (1999) 005, {\tt arXiv:hep-th/9908044};\\
A.\,Maloney, M.\,Spradlin, A.\,Strominger,
{\it Superconformal multi-black hole moduli spaces in four dimensions},
JHEP {\bf 0204} (2002) 003, {\tt arXiv:hep-th/9911001};\\
R.\,Britto-Pacumio, J.\,Michelson, A.\,Strominger, A.\,Volovich,
{\it Lectures on superconformal quantum mechanics and multi-black hole
moduli spaces}, published in ``Cargese 1999,
Progress in string theory and M-theory'', 235-264, {\tt arXiv:hep-th/9911066};\\
G.\,Papadopoulos, {\it Conformal and superconformal mechanics},
Class. Quant. Grav. {\bf 17} (2000) 3715, {\tt arXiv:hep-th/0002007}.

\bibitem{AFF}
V.\,de\,Alfaro, S.\,Fubini, G.\,Furlan,
{\it Conformal invariance in quantum mechanics},
Nuovo Cim. {\bf A34} (1976) 569.

\bibitem{FM}
D.Z.\,Freedman, P.F.\,Mende, {\it  An Exactly Solvable N Particle System In Supersymmetric
Quantum Mechanics}, Nucl. Phys. {\bf B344} (1990) 317.

\bibitem{Wyl}
N.\,Wyllard,
{\it (Super)conformal many body quantum mechanics with extended supersymmetry},
J. Math. Phys. {\bf 41} (2000) 2826, {\tt arXiv:hep-th/9910160}.

\bibitem{Vas}
L.\,Brink, T.H.\,Hansson, M.A.\,Vasiliev, {\it Explicit solution to the N body Calogero
problem}, Phys. Lett. {\bf B286} (1992) 109, {\tt arXiv:hep-th/9206049};\\
L.\,Brink, T.H.\,Hansson, S.\,Konstein, M.A.\,Vasiliev, {\it Anyonic representation, fermionic
extension and supersymmetry}, Nucl. Phys. {\bf B401} (1993) 591, {\tt arXiv:hep-th/9302023}.

\bibitem{GLP}
A.\,Galajinsky, O.\,Lechtenfeld, K.\,Polovnikov, {\it  Calogero models and nonlocal conformal
transformations}, Phys. Lett. {\bf B643} (2006) 221, {\tt arXiv:hep-th/0607215}; {\it
 N=4 superconformal Calogero models}, JHEP {\bf 0711} (2007) 008, {\tt
arXiv:0708.1075\,[hep-th]}; {\it N=4 mechanics, WDVV equations and roots}, JHEP {\bf 0903} (2009)
113, {\tt arXiv:0802.4386\,[hep-th]}.

\bibitem{FIL-09}
S.\,Fedoruk, E.\,Ivanov, O.\,Lechtenfeld, {\it OSp(4$|$2)
superconformal mechanics}, JHEP {\bf 0908} (2009) 081, {\tt
arXiv:0905.4951\,[hep-th]}.

\bibitem{KL-09}
S.\,Krivonos, O.\,Lechtenfeld,
{\it SU(2) reduction in N=4 supersymmetric mechanics},
Phys. Rev. {\bf D80} (2009) 04501, {\tt arXiv:0906.2469 [hep-th]}.

\bibitem{IKL}
E.A.\,Ivanov, S.O.\,Krivonos, V.M.\,Leviant,
{\it Geometric superfield approach to superconformal mechanics},
J. Phys. {\bf A22} (1989) 4201.

\bibitem{IL}
E.\,Ivanov, O.\,Lechtenfeld,
{\it N=4 supersymmetric mechanics in harmonic superspace},
JHEP {\bf 0309} (2003) 073, {\tt arXiv:hep-th/0307111}.

\bibitem{DI1}
F.\,Delduc, E.\,Ivanov,
{\it Gauging N=4 supersymmetric mechanics II:
(1,4,3) models from the (4,4,0) ones},
Nucl. Phys. {\bf B770} (2007) 179, {\tt arXiv:hep-th/0611247}.

\bibitem{IKLech}
E.\,Ivanov, S.\,Krivonos, O.\,Lechtenfeld,
{\it New variant of N=4 superconformal mechanics},
JHEP {\bf 0303} (2003) 014, {\tt arXiv:hep-th/0212303}.

\bibitem{IKLecht}
E.\,Ivanov, S.\,Krivonos, O.\,Lechtenfeld,
{\it N=4, d=1 supermultiplets from nonlinear realizations of $D(2,1;\alpha)$},
Class. Quant. Grav. {\bf 21} (2004) 1031, {\tt arXiv:hep-th/0310299}.

\bibitem{FJ}
L.\,Faddeev, R.\,Jackiw,
{\it Hamiltonian Reduction of Unconstrained and Constrained Systems},
Phys. Rev. Lett. {\bf 60} (1988) 1692;\\
G.V.\,Dunne, R.\,Jackiw, C.A.\,Trugenberger,
{\it ``Topological'' (Chern-Simons) quantum
mechanics}, Phys. Rev. {\bf D41} (1990) 661.

\bibitem{RPS}
R.\,Floreanini, R.\,Percacci, E.\,Sezgin, {\it Sigma Models With Purely Wess-Zumino-Witten
Actions}, Nucl. Phys. {\bf B322} (1989) 255.

\bibitem{HTown}
P.S.\,Howe, P.K.\,Townsend,
{\it Chern-Simons Quantum Mechanics}, Class. Quant. Grav. {\bf 7}
(1990) 1655.

\bibitem{Poly}
A.P.\,Polychronakos, {\it Integrable systems from gauged matrix models}, Phys. Lett. {\bf B266} (1991) 29.

\bibitem{IM}
E.\,Ivanov, L.\,Mezincescu, P.K.\,Townsend, {\it Fuzzy $CP(n|m)$ as a quantum superspace},
Contribution to "Symmetries in Gravity and Field Theory",
conference for Jose-Adolfo de Azcarraga's 60th birthday, June 2003, Salamanca, Spain,
{\tt arXiv:hep-th/0311159};\\
L.\,Mezincescu, {\it Super Chern-Simons Quantum Mechanics},
Proceedings of the International Workshop "Supersymmetries and Quantum Symmetries"
(SQS'03, 24-29 July, 2003), Dubna, Russia,
{\tt arXiv:hep-th/0405031}.

\bibitem{Sorba}
L.\,Frappat, A.\,Sciarrino, P.\,Sorba,
{\it Dictionary on Lie Algebras and Superalgebras},
Academic Press, 2000, {\tt arXiv:hep-th/9607161}.

\bibitem{BILS}
I.A.\,Bandos, E.\,Ivanov, J.\,Lukierski, D.\,Sorokin,
{\it On the superconformal flatness of AdS superspaces},
JHEP {\bf 0206} (2002) 040, {\tt arXiv:hep-th/0205104}.

\bibitem{HKLN}
T.\,Hakobyan, S.\,Krivonos, O.\,Lechtenfeld, A.\,Nersessian,
{\it Hidden symmetries of integrable conformal mechanical systems},
Phys. Lett. {\bf A374} (2010) 801, {\tt arXiv:0908.3290 [hep-th]}.

\bibitem{IN}
E.\,Ivanov, J.\,Niederle, {\it Biharmonic Superspace for N=4 Mechanics}, Phys. Rev. {\bf D80} (2009)
065027, {\tt arXiv:0905.3770\,[hep-th]}.

\bibitem{Je}
J.\,Van\,der\,Jeugt,
{\it Irreducible representations of the exceptional Lie superalgebras
$D(2,1;\alpha)$},
J. Math. Phys. {\bf 26} (1985) 913.

\bibitem{Barg}
V.\,Bargmann, {\it Representations of the Lorentz group}, Ann. Math. {48}
(1947) 568.

\bibitem{Per}
A.M.\,Perelomov,
{\it Algebraical approach to the solution of one-dimensional model
of n interacting particles},
Teor. Mat. Fiz. {\bf 6} (1971) 364 (in Russian).

\bibitem{AIPT}
J.A.\,de\,Azcarraga, J.M.\,Izquierdo, J.C.\,Perez\,Bueno, P.K.\,Townsend,
{\it Superconfor\-mal mechanics and nonlinear realizations},
Phys. Rev. {\bf D59} (1999) 084015, {\tt arXiv:hep-th/9810230}.

\bibitem{ABS}
V.P.\,Akulov, I.A.\,Bandos, D.P.\,Sorokin,
{\it Particle in harmonic N=2 superspace},
Sov. J. Nucl. Phys. {\bf 47} (1988) 724 [Yad. Fiz. {\bf 47} (1988) 1136];
{\it Particle mechanics in harmonic superspace},
Mod. Phys. Lett. {\bf A3} (1988) 1633.

\bibitem{ikrpa}
E.A.\,Ivanov, S.O.\,Krivonos, A.I.\,Pashnev, {\it Partial supersymmetry breaking in N=4
supersymmetric quantum mechanics}, Clas. Quant. Grav. {\bf 8} (1990) 19.

\bibitem{GGHPR}
J.P.\,Gauntlett, J.B.\,Gutowski, C.M.\,Hull, S.\,Pakis, H.S.\,Reall,
{\it All supersymmetric solutions of minimal supergravity in five- dimensions},
Class. Quant. Grav. {\bf 20} (2003) 4587, {\tt arXiv:hep-th/0209114}.

\bibitem{GMT-2}
J.P.\,Gauntlett, R.C.\,Myers, P.K.\,Townsend, {\it Black holes of D = 5 supergravity},
Class. Quant. Grav. {\bf 16} (1999) 1, {\tt arXiv:hep-th/9810204}.

\bibitem{Per-b}
A.M.\,Perelomov, {\it Integrable Systems of Classical Mechanics and Lie Algebras},
Birkhauser Verlag, Boston, 1990, 307pp. [in Russian: Nauka, Fizmatlit, 1990, 240pp.].

\bibitem{BMPV}
J.C.\,Breckenridge, R.C.\,Myers, A.W.\,Peet, C.\,Vafa, {\it D-branes and spinning black
holes}, Phys. Lett. {\bf B391} (1997) 93,  {\tt arXiv:hep-th/9602065}.

\bibitem{Town}
P.K.\,Townsend, {\it Killing spinors, supersymmetries and rotating intersecting branes}, in
the Proceedings of 22nd Johns Hopkins Workshop on Novelties of String Theory, Goteborg,
Sweden, 20-22 Aug 1998, {\tt arXiv:hep-th/990110};
{\it Surprises with angular momentum}, Annales Henri Poincare {\bf 4} (2003) S183, {\tt
arXiv:hep-th/0211008}

\bibitem{GibR}
G.W.\,Gibbons, P.\,Rychenkova, {\it Cones, Tri-Sasakian Structures and Superconformal Invariance},
Phys. Lett. {\bf B443} (1998) 138, {\tt arXiv:hep-th/9809158}.

\bibitem{Cas}
R.\,Casalbuoni, {\it On the Quantization of Systems with Anticommutating Variables}, Nuovo
Cim. {\bf A33} (1976) 115, {\it The Classical Mechanics for Bose-Fermi Systems}, Nuovo
Cim. {\bf A33} (1976) 389.

\end{thebibliography}
\end{document}